\documentclass[conference]{IEEEtran}
\pdfoutput=1

\IEEEoverridecommandlockouts
\pdfoutput=1
\usepackage{cite}
\usepackage{amsmath,amssymb,amsfonts}
\usepackage{algorithm}
\usepackage{algorithmic}
\usepackage{authblk}
\usepackage{graphicx}
\usepackage{textcomp}
\usepackage{xcolor}
\usepackage{booktabs}
\usepackage{longtable}
\usepackage{makecell}
\usepackage{multirow}
\usepackage{subfig}
\usepackage{verbatim}
\usepackage{tikz}
\usepackage{mathrsfs}

\def\BibTeX{{\rm B\kern-.05em{\sc i\kern-.025em b}\kern-.08em
    T\kern-.1667em\lower.7ex\hbox{E}\kern-.125emX}}
\begin{document}

\title{
  DAPPLE: A Pipelined Data Parallel Approach for Training Large Models
}

\author[1]{ Shiqing Fan}
\author[1]{ Yi Rong}
\author[1]{ Chen Meng}
\author[1]{ Zongyan Cao}
\author[1]{ Siyu Wang}
\author[1]{ Zhen Zheng}
\author[2]{ \\Chuan Wu} 
\author[1]{ Guoping Long}
\author[1]{ Jun Yang}
\author[1]{ Lixue Xia}
\author[1]{ Lansong Diao}
\author[1]{ Xiaoyong Liu}
\author[1]{ Wei Lin}

\affil[1]{\footnotesize Alibaba Group, China}
\affil[2]{\footnotesize The University of Hong Kong, China}
\affil[ ]{\{shiqing.fsq, rongyi.ry, mc119496, zongyan.cao, siyu.wsy, james.zz\}@alibaba-inc.com}
\affil[ ]{cwu@cs.hku.hk, \{guopinglong.lgp, muzhuo.yj, lixue.xlx, lansong.dls, xiaoyong.liu, weilin.lw\}@alibaba-inc.com}

\maketitle

%-------------------------------------------------------------------------------
\begin{abstract}
%-------------------------------------------------------------------------------
It is a challenging task to train large DNN models on sophisticated
GPU platforms with diversified interconnect capabilities.
Recently, pipelined training has been proposed
as an effective approach for improving device utilization.
However, there are still several tricky issues to address: 
improving computing efficiency while ensuring convergence, and
reducing memory usage without incurring additional computing costs.
We propose \emph{DAPPLE}, a synchronous training framework which combines
data parallelism and pipeline parallelism for large DNN models.
It features a novel parallelization strategy \emph{planner} %for synchronous training(friendly for model convergence)
to solve the partition and placement problems, and explores the optimal hybrid strategies of data and pipeline parallelism.
We also propose a new runtime scheduling algorithm to reduce device
memory usage, which is orthogonal to re-computation approach and does not come
at the expense of training throughput.
Experiments show that \emph{DAPPLE planner} consistently outperforms strategies generated by PipeDream‘s planner by up to $3.23\times$ speedup under synchronous training scenarios, and \emph{DAPPLE runtime} outperforms GPipe by $1.6\times$ speedup of training throughput and saves 12\% of memory consumption at the same time.
%given a fixed global batch size, 
% \emph{DAPPLE} outperforms the best data parallelism baselines with 1.71X/1.37X/1.79X % (up to 2.32X for GNMT-16% on \cb{config $C$})
% training speedups on three typical cluster environments.
% Note: these number is calculated when **GBS=128** for all models

\end{abstract}

\begin{IEEEkeywords}
deep learning, data parallelism, pipeline parallelism, hybrid parallelism
\end{IEEEkeywords}

\section{Introduction}
\label{sec:intro}

%Deep Neural Networks (DNNs) have achieve great success in
%many fields, including image classification \cite{krizhevsky2012imagenet}
%\cite{simonyan2014very},
%language model \cite{merity2017regularizing},
%machine translation \cite{bahdanau2014neural} \cite{arivazhagan2019massively}, etc.
%Existing deep learning frameworks such as Tensorflow \cite{tensorflow},
%PyTorch \cite{pytorch} and Caffe2 \cite{caffe2}
%\dots
%In recent years, 
The artificial intelligence research community has a long history of harnessing computing power 
to achieve significant breakthroughs \cite{bitterlesson}. For deep learning, a 
trend has been increasing the model scale up to the limit of modern AI hardware. 
Many state-of-the-art DNN models (e.g., NLP\cite{t5}, 
Internet scale E-commerce search/recommendation systems\cite{jizhe, youtube}) have 
billions of parameters, demanding tens to hundreds of GBs of device memory for training. A critical 
challenge is how to train such large DNN models on hardware accelerators, 
such as GPUs, with diversified interconnect capabilities.

%A common approach for DNN training is data parallelism (\emph{DP}). With \emph{DP} training,
A common approach is sychronous data parallel (\emph{DP}) training.
Multiple workers each performs complete model computation and
synchronizes gradients periodically to ensure proper model convergence.
\emph{DP} is simple to implement and friendly in terms of load balance, but 
%its feasibility and performance depend on the model scale and the computation/communication ratio. 
%the performance depends on the computation/communication ratio.
%Parameter communication overhead is a major factor preventing linear scalability of \emph{DP}, which can be 
the gradients sychronization overhead can be a major factor preventing linear scalability.
While the performance issue can be
alleviated by optimizations such as local gradients accumulation\cite{GA-TF, GA-Torch, caffe2}
or computation and communication overlap techniques\cite{jayarajan2019priority, sergeev2018horovod},
aggressive \emph{DP} typically requires large training batch sizes, which makes model tuning harder. 
More importantly, \emph{DP} is
not feasible once the parameter scale of the model exceeds the memory limit of a single device.

Recently, pipeline parallelism\cite{huang2019gpipe,zhan2019pipe,narayanan2019pipedream}
has been proposed as a promising approach for training large DNN models.
The idea is to partition model layers into multiple groups (stages) and place them on a set of
inter-connected devices.
During training, each input batch is further divided into multiple micro-batches, which are scheduled
to run over multiple devices in a pipelined manner.
Prior research on pipeline training generally falls into two categories.
One is on optimizing pipeline parallelism for synchronous training\cite{huang2019gpipe,zhan2019pipe}.
This approach requires necessary gradients synchronizations between adjacent
training iterations to ensure convergence. At runtime, it schedules as many concurrent pipe stages
as possible in order to maximize device utilization. In practice, this scheduling policy can
incur notable peak memory consumption. To remedy this issue, re-computation\cite{recomputation} 
can be introduced to trade redundant computation costs for reduced memory usage.
Re-computation works by checkpointing nodes in the computation graph defined by user model,
and re-computing the parts of the graph in between those nodes during backpropagation.
%It divides a mini-batch into multiple micro-batches and feeds them into pipeline to improve device
%utilization. This manner inserts synchronizations between different pipeline iterations to keep the 
%original training logic. 
The other category is asynchronous(\emph{async}) pipeline training 
\cite{narayanan2019pipedream}. 
This manner inserts mini-batches into pipeline continuously 
and discards the original sync operations to achieve maximum throughput.

Although these efforts have made good contributions to advance pipelined training techniques, 
they have some serious limitations. 
%\cb{Although \cite{narayanan2019pipedream} 
%made good progress to improve the time-to-accuracy(TTA) for some models with asynchronous pipeline training mechanism, 
%due to the diversity and fast-evolution of model architectures, async training is not a common practice in 
%important industry application domains for convergence concern, which is also reflected in \cite{wang2019characterizing}.}
While \emph{PipeDream}\cite{harlap2018pipedream} made progress in improving the time-to-accuracy for some benchmarks with \emph{async}
pipeline parallelism, \emph{async} training is not a common practice in important industry application domains due to 
convergence concerns. This is reflected in a characterization study\cite{wang2019characterizing} 
of widely diversified and fast evolving workloads in industry scale clusters.
In addition, the \emph{async} approach requires 
the storage of multiple versions of model parameters. This, while friendly for increasing parallelism, 
further exacerbates the already critical memory consumption issue. As for synchronous training,
current approach\cite{huang2019gpipe} still requires notable memory consumption, because no backward processing(BW) can be scheduled 
until the forward processing(FW) of all micro-batches is finished. The intermediate results of some micro-batches in FW 
need to be stored in the memory (for corresponding BW's usage later) while the devices are busy with FW of some other micro-batches.
GPipe\cite{huang2019gpipe} proposes to discard some intermediate results to free the memory and 
re-computes them during BW when needed. But this may introduce additional $\sim20\%$
re-computation overhead\cite{gpipeTalk19}. %% TODO@shiqing: update.

% \begin{table}[t]
%     \caption{Important notations used in this paper}
%     \label{tbl:notations}
%     \begin{center}
%     \begin{tabular}{ll}
%     \toprule
%     Notation & Description \\
%     \midrule
%     $S$ & Number of stages \\ %(indexed from $1$)
%     $M$ & Number of equal micro-batches \\
%     $\alpha$ & Communication-to-computation ratio \\
%     $N$ & Number of layers in model \\
%     \bottomrule
%     \end{tabular}
%     \end{center}
%   \end{table}
  
In this paper, we propose \emph{DAPPLE}, a distributed training scheme which combines pipeline parallelism and data parallelism to 
ensure both training convergence and memory efficiency.
\emph{DAPPLE} adopts synchronous training
to guarantee convergence, while avoiding the storage of multiple versions of parameters in \emph{async} approach.
Specifically, we address two design challenges.
The first challenge is how to determine an optimal parallelization strategy given model structure and 
hardware configurations. The \emph{optimal} strategy refers to the one where the execution time of a single global step theoretically reaches the minimum for given resources. The target optimization space includes \emph{DP}, pipelined parallelism, and hybrid approaches combining both. Current state-of-the-art pipeline partitioning algorithm \cite{narayanan2019pipedream}
is not able to be applied to synchronous training effectively.
Some other work\cite{huang2019gpipe,gpipeTalk19} relies on empirical and manual optimizations, 
and still lacks consideration of some parallelism dimensions.
We introduce a synchronous pipeline planner, which generates optimal parallelization strategies automatically
by minimizing execution time of training iterations.
Our planner combines pipeline and data parallelism (via stage-level replication) together while partitioning layers into
multiple stages. Besides pipeline planning, for those models that can fit into % to Problem-15
a single device and with high computation/communication ratio, 
the planner is also capable of producing \emph{DP} strategies directly for runtime execution.
Furthermore, the planner takes both memory usage and interconnect topology constraints
into consideration when assigning pipeline stages to devices.
The assignment of computation tasks to devices is critical for distributed training performance in hardware environments with complex interconnect configurations. In \emph{DAPPLE}, three topology-aware device assignment mechanisms are defined and incorporated into the pipeline
partitioning algorithm.

The second challenge is how to schedule pipeline stage computations, in order to achieve a balance among
parallelism, memory consumption and execution efficiency.
We introduce \emph{DAPPLE} schedule, a novel pipeline stage scheduling algorithm which achieves
decent execution efficiency with reasonably low peak memory consumption.
% thus avoiding re-computation overheads completely.
A key feature of our algorithm is to schedule forward and backward stages in a deterministic and interleaved manner to 
release the memory of each pipeline task as early as possible.

%\TODO{Experiments show that \emph{DAPPLE} significantly boosts the efficiency of ... Data show
%It does not only work on large model, but also work on model like VGG (brief describe that it reduces 
%the communication overhead)}
We evaluate \emph{DAPPLE} on three representative application domains, namely image classification, 
machine translation and language modeling. For all benchmarks, 
experiments show that our planner can consistently
produce optimal hybrid parallelization strategies combining data and pipeline parallelism on 
three typical GPU hardware environments in industry, i.e. hierarchical \emph{NVLink + Ethernet},
$25$ Gbps and $10$ Gbps \emph{Ethernet} interconnects.
Besides large models, \emph{DAPPLE} also works well for medium scale models with relatively 
large weights yet small activations (i.e. VGG-19). Performance results show that (1) \emph{DAPPLE} 
can find optimal hybrid parallelization strategy outperforming the best \emph{DP} baseline up to $2.32\times$ speedup; (2) \emph{DAPPLE planner} consistently outperforms strategies generated by PipeDream‘s planner by up to $3.23\times$ speedup under synchronous training scenarios, and (3) \emph{DAPPLE runtime} outperforms GPipe by $1.6\times$ speedup of training throughput and reduces the memory consumption of 12\% at the same time.

The contributions of \emph{DAPPLE} are summarized as follows:

\begin{enumerate}

\item We systematically explore hybrid of data and pipeline parallelism
with a pipeline stage partition algorithm for \emph{synchronous}
training, incorporating a topology-aware device assignment
mechanism given model graphs and hardware configurations. This facilitates large model training and reduces communication overhead of \emph{sync} training, which is friendly for model convergence.

\item We feature a novel parallelization strategy \emph{DAPPLE planner} to solve the partition and placement problems and explore the optimal hybrid strategies of data and pipeline parallelism, which consistently outperforms SOTA planner's strategies under synchronous training scenarios.

\item We eliminate the need of storing multiple versions of parameters, \emph{DAPPLE} introduces a pipeline task scheduling approach to further reduce memory consumption. This method is $orthogonal$ to re-computation approach and does not come at the expense of training throughput. Experiments show that \emph{DAPPLE} can further save about 20\% of device memory on the basis of enabling re-computation optimization.

\item We provide a \emph{DAPPLE} runtime which realizes efficient 
pipelined model training with above techniques without compromising model convergence accuracy.

\end{enumerate}

\section{Motivation and \emph{DAPPLE} Overview}
\label{section:overview}
We consider pipelines training only if \emph{DP} optimizations 
\cite{goyal2017accurate,FT1,FT2,poseidon,jayarajan2019priority,GA-Torch,GA-TF}
are unable to achieve satisfactory efficiency, or the model size is too large to fit in a device with a minimum required batch size. In this section, we summarize key design issues in synchronous training with parallelism and motivate our work.

\subsection{Synchronous Pipeline Training Efficiency}

Pipeline training introduces explicit data dependencies between consecutive stages (devices). A common approach to keep all stages busy is to split the training batch into multiple micro-batches\cite{huang2019gpipe}. These micro-batches are scheduled in the pipelined manner to be executed on different devices concurrently.%, as shown in Figure \ref{fig:gpipe-schedule}.
Note that activation communication
(\emph{comm}) overhead matters in synchronous training.
Here we incorporate \emph{comm} as a special pipeline stage for our analysis.
We define \emph{pipeline efficiency} as average GPU utilization of all devices in the pipeline. The pipeline efficiency is $\frac{1}{1+P}$\cite{huang2019gpipe}, 
where $P=\frac{(1+\alpha)\dot(S-1)}{M}$. $S$, $M$ and $\alpha$ are number of stages, number of equal micro-batches and communication-to-computation ratio, respectively. Given a pipeline partitioning scheme (namely fixed $S$ and $\alpha$), the larger the number of micro-batches $M$, the higher the pipeline efficiency. Similarly, the smaller $S$, the higher the efficiency  for fixed $\alpha$ and $M$,.

\begin{comment}
\begin{figure}
    \includegraphics[width=0.4\textwidth]{./background/figs/gpipe_network.png}
    \caption{
    Synchronous Scheduling
    }
    \label{fig:gpipe-schedule}
\end{figure}
\end{comment}

There are efforts to improve synchronous pipelines with micro-batch scheduling \cite{huang2019gpipe}, which suffer from two issues. 

(1) Extra memory consumption. State-of-the-art approach injects all micro-batches consecutively into the first stage of FW. Then the computed activations are the input to BW. Execution states (i.e. activations) have to be kept in memory for all micro-batches until their corresponding BW operations start. Therefore, while more injected micro-batches may imply higher efficiency, the memory limit throttles the number of micro-batches allowed.

(2) Redundant computations. State-of-the-art approach may adopt activation re-computation to reduce peak memory consumption \cite{recomputation}, i.e. discarding some activations during the FW phase, and recomputing them in the BW phase when needed. However, redundant computations come with extra costs. It is reported that re-computation can consume approximately 20\% more execution time\cite{gpipeTalk19}.

\subsection{Pipeline Planning}
To maximize resource utilization and training throughput for pipelined training, it is crucial to find a good strategy for partitioning stages and mapping stages to devices. We refer to the stage partitioning and device mapping problem as \emph{pipeline planning problem}. Current state-of-the-art planning algorithm \cite{harlap2018pipedream} may suffer from the following issues.

First, it does not consider synchronous pipeline training. Synchronous pipeline is very important as convergence is the prerequisite of training. Compared against \emph{async} pipeline training, an additional step is needed for \emph{sync} pipeline training at the end of all micro-batches to synchronize parameter updates. In more generic case where a stage may be replicated on multiple devices, there exists additional gradients synchronizations (\emph{AllReduce}) overheads before parameter updates.
Current pipeline planner does not take such overhead into consideration 
and thus can not accurately model the end-to-end pipeline training time. 

Second, previous approach does not consider the impact of the number of stages $S$,
which is important for synchronous planning. As discussed in the previous subsection, with fixed micro-batches $M$ and \emph{comm} overhead ratio $\alpha$, the fewer the number of stages($S$) , the higher the pipeline efficiency.

Finally, \emph{uneven} stage partitions have not been well studied in existing works. We show in Section \ref{sec:planner} that \emph{uneven} stage partitions can sometimes produce better training performance.

\subsection{The DAPPLE Approach}
\begin{figure}[t]
    \includegraphics[width=0.45\textwidth]{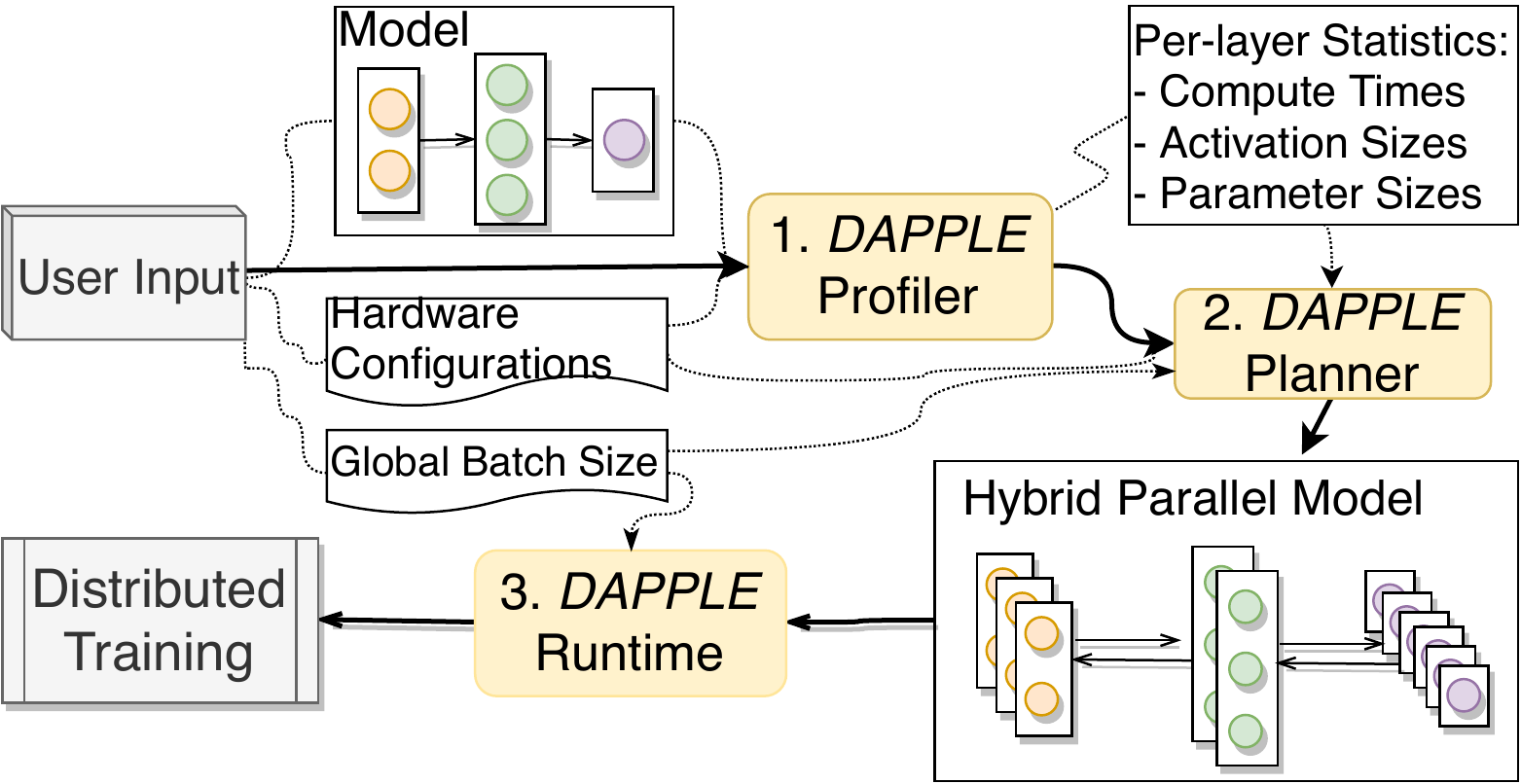}
    \caption{
    \emph{DAPPLE} framework overview.
    }
    \label{fig:overview}
\end{figure}

We propose the \emph{DAPPLE} framework to address aforementioned scheduling and planning challenges with synchronous training.
Fig. \ref{fig:overview} shows high-level workflow of \emph{DAPPLE}. It features a \emph{profiler}, a \emph{planner} and a \emph{runtime} system. Overall, \emph{DAPPLE profiler} takes a user's DNN model as input, and profiles execution time, activation sizes and parameter sizes for each layer. Taking profiling results as input, \emph{DAPPLE planner} generates an optimized (hybrid) parallelization plan on a given \emph{global batch size}. In terms of the execution time, both \emph{DAPPLE profiler} and \emph{planner} are offline and can be completed within a few seconds for all our benckmark models (Table \ref{tbl:models}). Finally \emph{DAPPLE runtime} takes the \emph{planner}'s results, and transforms the original model graph into a pipelined parallel graph. At this stage, \emph{global batch size} is further split into multiple micro-batches and then been scheduled for execution by \emph{DAPPLE runtime}. 

More specifically, the \emph{planner} aims at minimizing the end-to-end 
execution time of one training iteration.
This module is responsible for stage partition, stage replication and device assignment
and generates optimal parallelization plans.
In particular, the device assignment process is aware of the hardware topology,
as shown in Fig. \ref{fig:overview}.

% \vspace{-1ex}
\begin{table}[t]
   \begin{small}
   \caption{The traffic volume. 
%    \textit{Activations} means the activation size between layers that \emph{DAPPLE} partitions. 
%    \textit{Gradients} is the overall gradients traffic with data parallel.
   }
   \label{tbl:comm-vol}
   \begin{center}
   \begin{tabular}{lcc}
   \toprule
   Benchmark & \makecell[c]{Activation Size at \\the Partition Boundaries}  & Gradient Size \\
   \midrule
   % Note: the pipeline comm volume needs to be multiplied by 2 for fw as well as bw phase.
   % Need to be double checked by @RONGYI
   GNMT-16 & $26$MB & $1.1$GB \\ % act: 0.4MB/sample, batch_size=32
   % GNMT-36 & $17.8$MB & $1.8$GB \\ % act: 0.139MB/sample, batch_size=32, ?:? staging
   BERT-48 & $8.8$MB & $2.8$GB \\ % act: 2.2 MB/sample, batch_size=2, ?:? staging
   XLNET-36 & $4.2$MB & $2.1$GB \\ % act: 2MB/sample, batch_size=1, ?:? staging
   % AmoebaNet-18 & $77.1$MB & $1.9$GB \\ % batch_size=8, 13:5 staging
   AmoebaNet-36 & $11.2$MB & $3.7$GB \\ % batch_size=1, 24:12 staging
   VGG-19 & $6$MB & $550$MB \\ % batch_size=32, 15:1 staging
   \bottomrule
   \end{tabular}
   \end{center}
   \end{small}
\end{table}

\begin{figure}[t]
    \includegraphics[width=0.45\textwidth]{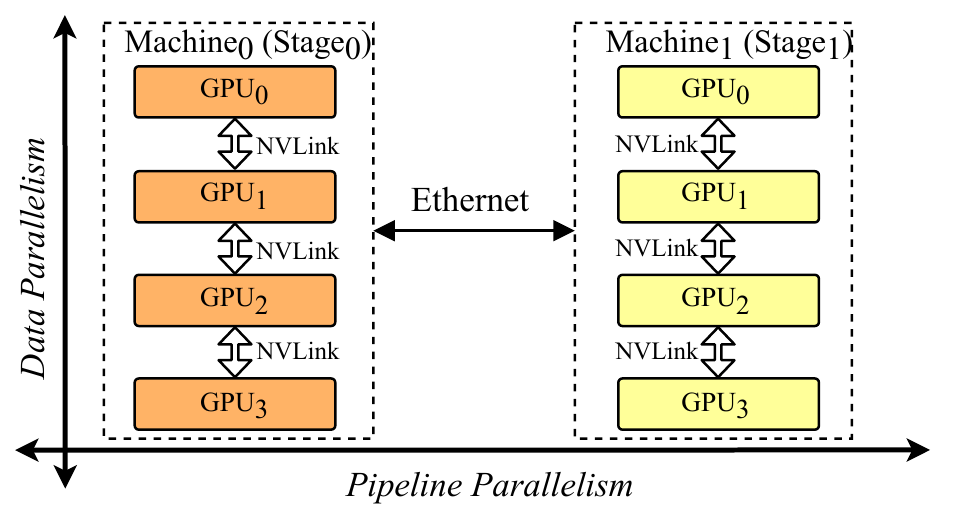}
    \vspace{-1ex}
    \caption{
    Device mapping on hierarchical interconnects.
    }
    \label{fig:dapple}
\end{figure}

We also explore the mapping of a single stage onto multiple devices. With the replication of pipeline stages on multiple devices, \emph{DAPPLE} processes training with the hybrid of 
data and pipeline parallelism. In practice, this hybrid strategy can exploit hierarchical interconnects effectively.
Fig. \ref{fig:dapple} gives an example where a model is partitioned into two stages and each stage is replicated on four
devices within the same server(NVLink connections within server),  while inter-stage communication goes over 
the Ethernet. This mapping exploits
workload characteristics (Table \ref{tbl:comm-vol})
by leveraging the high-speed NVLink for 
heavy gradients sync, while using the slow Ethernet bandwidth for small activations 
communication. We discuss details of our planner in Section \ref{sec:planner}.

Finally, the \emph{DAPPLE runtime} involves a carefully designed scheduling algorithm.
Unlike existing pipeline scheduling algorithms \cite{yang2019pipemare}, 
\emph{DAPPLE} scheduler (Section \ref{section:scheduler}) 
significantly reduces the need for re-computation, retains a reasonable
level of memory consumption, while saturates the pipeline with enough 
micro-batches to keep all devices busy.

% \vspace{-2ex}
% \vspace{-2ex}

\section{\emph{DAPPLE} Schedule}
\label{section:scheduler}

\subsection{Limitations of GPipe Schedule}
\label{limitation-of-gpipe}
\vspace{-0.5ex}

\begin{figure}[t]
    \includegraphics[width=\linewidth] {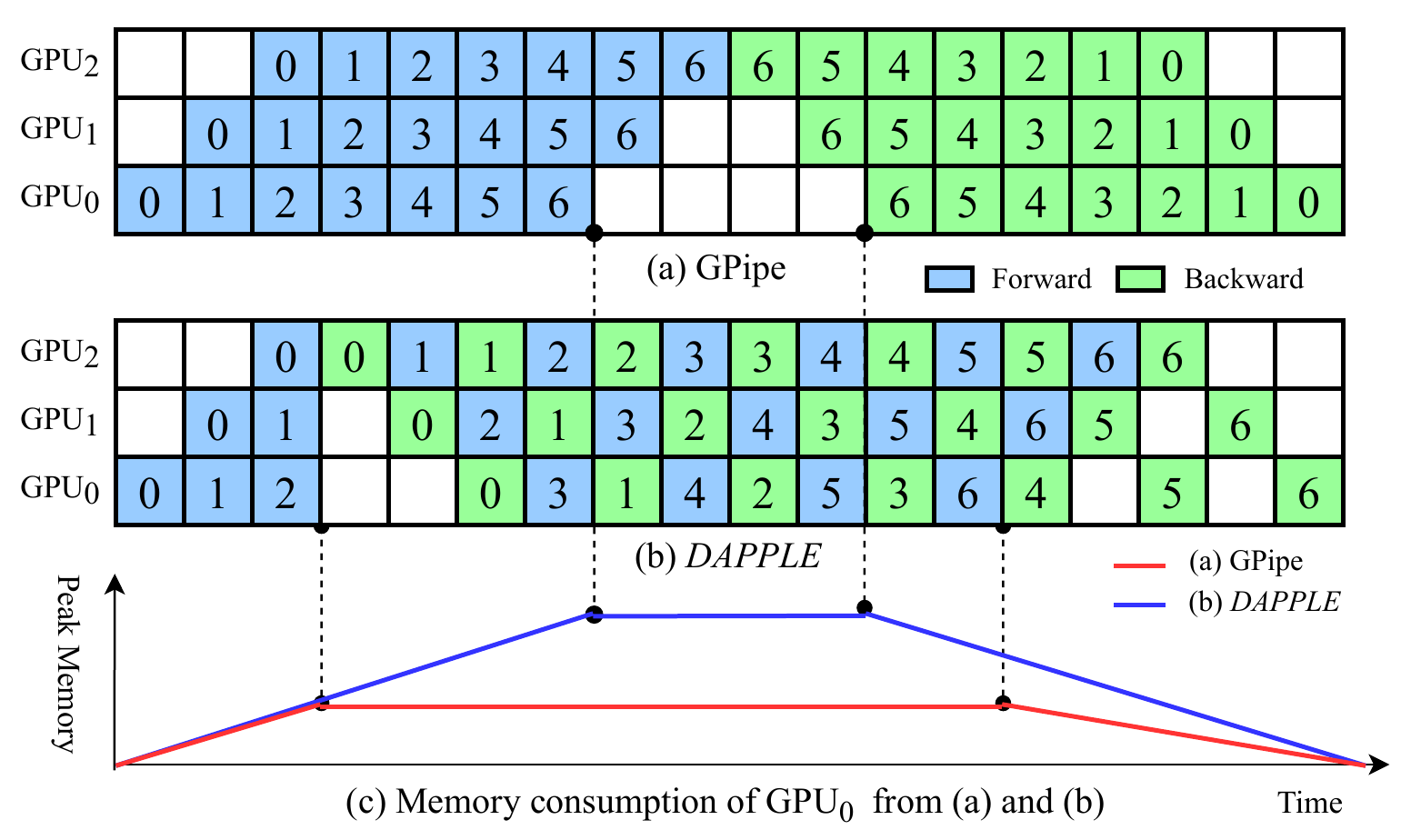}
    % \vspace{-2ex}
    \caption{
        The different scheduling between GPipe(a) and DAPPLE(b) and their memory consumptions.
    }
    \label{fig:gpipe_dapple}
\end{figure}

To improve pipeline training efficiency, GPipe\cite{huang2019gpipe} proposes to split 
global batch into multiple micro-batches and injects them into the pipeline concurrently (Fig.
\ref{fig:gpipe_dapple} (a)). However, this scheduling pattern alone is not
memory-friendly and will not scale well with large batch.
The activations produced by forward tasks have to be kept for all micro-batches
until corresponding backward tasks start, thus leads to the memory demand to be proportional ($O(M)$)
to the number of concurrently scheduled micro-batches ($M$). GPipe adopts
\textit{re-computation} to save memory while brings approximately 20\% extra computation.
In \emph{DAPPLE}, we propose \textit{early backward scheduling} to reduce memory consumptions 
while achieving good pipeline training efficiency(Fig. \ref{fig:gpipe_dapple} (b)).

\vspace{-0.5ex}
{\subsection{Early backward scheduling}
\label{early-backward-scheduling}}
\vspace{-0.5ex}

The main idea is to schedule backward tasks(BW) earlier and hence free the memory used for storing
activations produced by corresponding forward tasks(FW). Fig. \ref{fig:gpipe_dapple}(b) shows 
\emph{DAPPLE}'s scheduling mechanism, compared to GPipe in Fig. \ref{fig:gpipe_dapple} (a). 
Firstly, instead of injecting all $M$ micro-batches at once, we propose to inject K micro-batches 
($K<M$) at the beginning to release memory pressure while retaining high pipeline efficiency. 
Secondly, we schedule one FW of a micro-batch followed by one BW strictly 
to guarantee that BW can be scheduled earlier. 
% In order to showcase our advantage in memory savings, we focus on the memory consumptions in
% the first stage since it consumes the most memory among all the stages. 
Fig. \ref{fig:gpipe_dapple} (c) shows
how the memory consumptions change over time in GPipe and \emph{DAPPLE}.
At the beginning, the memory usage in \emph{DAPPLE} increases with time and
is the same as GPipe's until $K$ micro-batches are injected, then it reaches the maximum due to the early BW scheduling.
Specifically, with strictly controlling the execution order of FW and BW, the occupied memory 
for activations produced by the FW of a micro-batch will be freed after the corresponding BW
so that it can be reused by the next injected micro-batch.
In comparison, GPipe's peak memory consumptions increases continuously and has no opportunity for
early release. Moreover, \emph{DAPPLE} does not sacrifice in pipeline training efficiency. Actually, 
\emph{DAPPLE} introduces the exact same bubble time as GPipe when given the same stage partition, micro-batches
and device mapping. We will present the details in section~\ref{section:micro-batch-schedule}.

Note that the combination of early
backward scheduling and re-combination allows further exploitation in memory usage.
We present performance comparisons of \emph{DAPPLE} and GPipe in section~\ref{sec:scheduling-policy}.

% \vspace{-2ex}

\vspace{-1ex}
\section{\emph{DAPPLE} Planner}
\label{sec:planner}
\vspace{-1ex}

\emph{DAPPLE Planner} generates an optimal hybrid parallelism execution plan 
given profiling results of \emph{DAPPLE profiler}, hardware configurations 
and a global training batch size.

% \vspace{-.5ex}
\begin{figure}[t]
  \centering
  \includegraphics[width=1.0\linewidth]{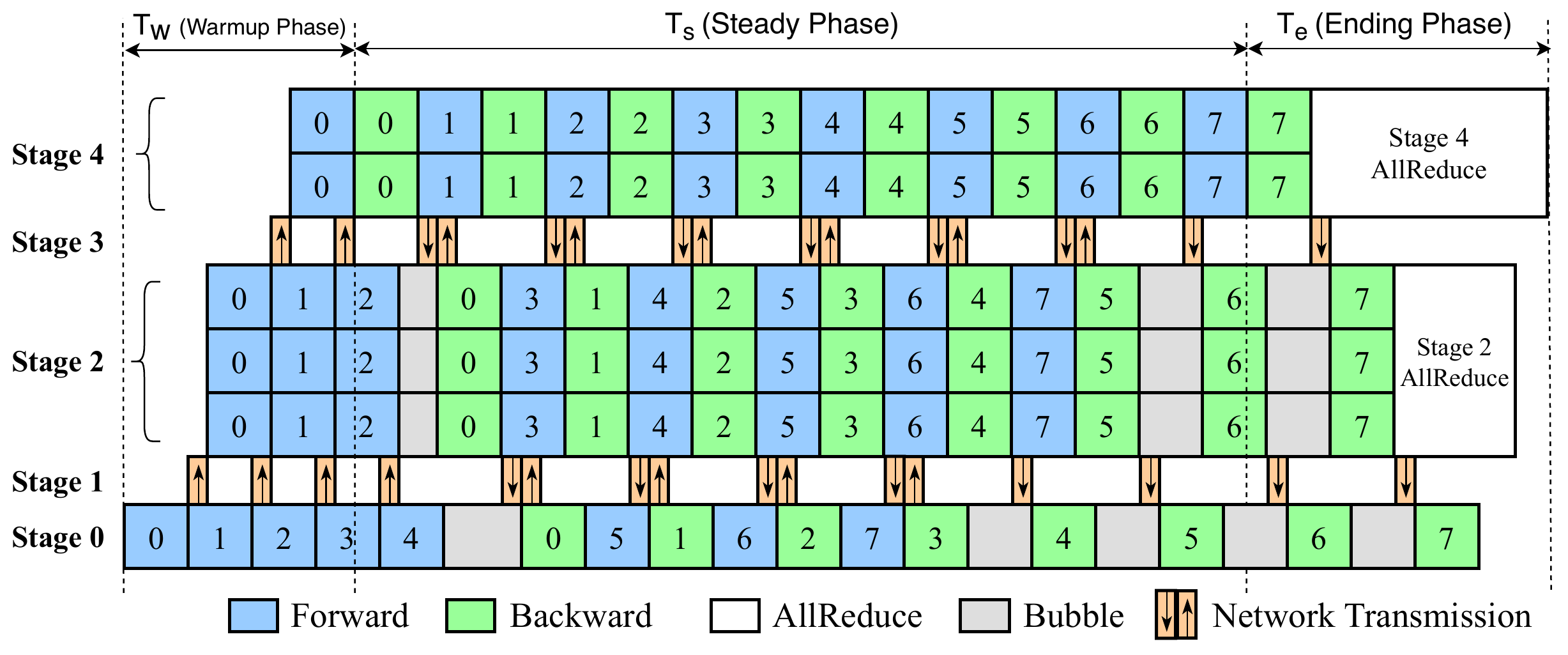}
  \vspace{-.5ex}
  \caption{\emph{DAPPLE} pipeline example.}
  \label{fig:DAPPLE-Planning-Pipeline}
\end{figure}
% \vspace{-.5ex}

\vspace{-.5ex}
\subsection{The Optimization Objective}
\label{sec:planner-objective}
\vspace{-.5ex}

% While previous works \cite{narayanan2019pipedream,torchgpipe,barany2015block}
% focus on balancing the workloads across all GPUs, we found out that focusing
% purely on evenness does not always result in better performance. \TODO{explain
%   why evenness is bad} \TODO{Or we just remove this paragraph}
% , details of
% which can be found in the Appendix \ref{sec:planner-result-insights}.

For synchronous training, we use the execution time of a single global batch
as our performance metric, which we call pipeline latency.
The optimization objective is to minimize pipeline latency \(L\)
with the consideration of all solution spaces of data/pipeline parallelism.

In synchronous pipelined training, computations and cross-stage communication of
all stages usually form a trapezoid, but not diamond formed by normal pipelines without backward phase.
Fig.\ref{fig:DAPPLE-Planning-Pipeline} shows a pipelined training example with 
well designed task scheduling arrangement. We use blue blocks for forward
computation, and green ones for backwards, with numbers in them being the
corresponding micro-batch index. Network communications are arranged as individual stages.
Gray blocks are bubbles.
% Such shape shown above is formed due to the nature of DNN
%training: for example, the forward block of micro-batch $i$ at stage $s$ must be
%performed before the forward block of micro-batch $i$ at stage $s+2$ as well as
%the forward block of micro-batch $i+1$ at stage $s$.

We denote the stage with the least bubble overhead as \textit{pivot
  stage}, which will be the dominant factor in calculating pipeline
latency $L$. Let its stage id be $Q$. We will discuss about how to choose pivot stage later.

A pipeline training iteration consists of warmup phase, steady phase and ending phase, 
as shown in Fig. \ref{fig:DAPPLE-Planning-Pipeline} in which pivot stage is the last stage. 
Pivot stage dominates steady phase.
We call the execution period from the start to \textit{pivot stage}'s first micro-batch
as warmup phase in a pipeline iteration, the period from \textit{pivot stage}'s last 
micro-batch to the end as ending phase. Pipeline latency is the sum of these three phases. 
The optimization objective for estimating $L$ is as follows:

\vspace{-1.5ex}
\begin{equation}
	\begin{split}
    T_w & = \sum_{s=0}^{Q} F_s \\
    T_s & = (M-1) \times (F_Q + B_Q)  \\
    % T_e & = \max_{s=0}^{S-1} \Bigg( (s \ge Q ? -\sum_{a=Q}^s B_a : \sum_{a=s}^Q B_a) + AR(P_s, g_s) \Bigg) \\
    T_e & = \max_{s=0}^{S-1} (AR(P_s, g_s) + 
    \begin{cases}
      -\sum_{a=Q}^s B_a& s \ge Q\\
      \sum_{a=s}^Q B_a& s < Q
    \end{cases}) \\
  \end{split}
\end{equation}

\vspace{-.5ex}
\begin{equation}
	L = T_w + T_s + T_e
	\label{eq:latency-objective}
\end{equation}
\vspace{-.5ex}

% TODO(@RongYi): Refine equation (1).
$T_w$ denotes the execution time of warmup phase, which is the sum of forward
execution time of stages till $Q$ for one micro-batch. $T_s$ denotes the steady
phase, which includes both forward and backward time of stage $Q$ for all
micro-batches except for the one contributing to warmup and ending phase. $T_e$
corresponds to the ending phase. $T_e$ includes allreduce overhead and thus 
considers stages both before and after $Q$. Note that some stages before $Q$ may
contribute to $T_e$ with allreduce cost. $M$, $S$, $F_s$ and $B_s$ denote the
total number of micro-batches, the number of stages (computation stages +
network stages), forward and backward computation time of stage $s$,
respectively. $AR(P_s,g_s)$ represents the gradients synchronization
(\emph{AllReduce}) time for stage $s$, with its parameter set $P_s$ on the
device set $g_s$.

Note we consider inter-stage communication as an independent stage alongside the
computation stages. The AllReduce time $AR(P_s, g_s)$ is always 0 for communication stages. 
Moreover, we define $F_s$ and $B_s$ for a communication stages as its following 
forward and backward communication time.

In practice, synchronous pipelines in some cases include bubbles in stage $Q$,
which may contribute a small fraction of additional delay to time.
This objective does not model those internal bubbles, and thus is an
approximation to the true pipeline latency.
But it works practically very well for all our benchmarks (Section \ref{section:evaluation}).

% The initial $Q$ is $S-1$. Formula \ref{eq:pivot-stage} along with Algorithm
% \ref{alg:iterative-q} describe how to update stage $Q$ iteratively from stage
% $S-1$ to stage $0$. $T_{st}^j = (M-1) \times (F_j + B_j)$ means the duration of
% steady phase, without bubbles, if pivot stage is $j$. For a stage $s<Q$, if
% $T_{st}^s$ is larger than the sum of $T_{st}^Q$ and corresponding
% forward/backward costs between $s$ and current $Q$, it means the steady phase
% will have less bubble if pivot stage is set as $s$ other than current $Q$. $Q$
% will then be updated to $s$.

% We present more details of our modeling rationale in the Appendix
% \ref{sec:planner-pipeline-length-modeling}. We also analyze the empirical bound
% between our estimation of $L$ and the ground truth in the Appendix
% \ref{sec:planner-empirical-analysis}.

\vspace{-.5ex}
\subsection{Device Assignment}
\label{subsec:device_assignment}
\vspace{-.5ex}

Device assignment affects communication efficiency and computing resource
utilization.
% The placement of each stage on a given set of GPUs with different
% interconnection bandwidth is an NP-Hard problem since there are
% $$\prod_{i=0}^{S-1} {(G - \sum_{j=0}^{i-1} {r_j}) \choose r_i} =
% \prod_{i=0}^{S-1} \frac{(G - \sum_{j=0}^{i-1} {r_j})!}{r_i ! \times (G -
%   (\sum_{j=0}^{i-1} {r_j}) - r_i)!}$$ placement possibilities, with $S$ denoting
% the number of stages, $r_i$ being the replication factor for stage $i$, and $G$
% denoting the total number of GPUs to be placed on.
Previous work \cite{narayanan2019pipedream} uses hierarchical planning 
and works well for asynchronous training. 
However, it lacks consideration of synchronous pipeline training, in which
the latency of the whole pipeline, rather than of a single stage, matters to
overall performance. It cannot be used to efficiently estimate the whole pipeline latency. 
Meanwhile, it does not allow stages to be placed on arbitrary devices. 
Our approach essentially allows a specific stage to be mapped to any
set of devices, and therefore is able to handle more placement cases, at a
reasonable searching cost.

\begin{figure}[!tbp]
  \centering
  \includegraphics[width=\linewidth]{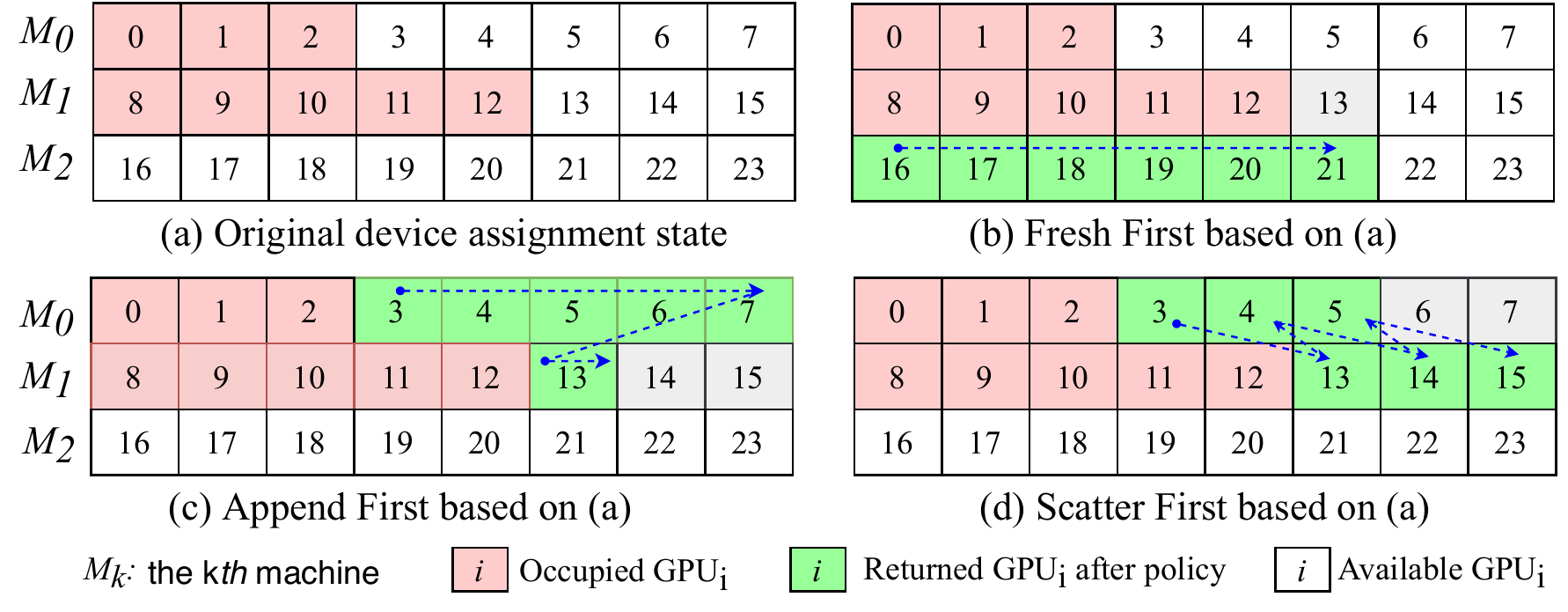}
  \caption{\label{fig:HPGO-Placement}
  Device assignment examples: applying for $6$
  devices using three different strategies respectively from (a).}
\end{figure}

Instead of enumerating all possibilities of placement plans using brute force,
we designed three policies (Fig. \ref{fig:HPGO-Placement}), and explore their compositions 
to form the final placement plan.

\paragraph{Fresh First} allocate GPUs from a fresh machine. 
It tends to put tasks within a stage onto the same machine,
which can leverage high-speed NVLink \cite{nvlink19} for intra-stage
communication.
A problem of this policy is that, it can cause fragmentation if the stage
cannot fully occupy the machine.

\paragraph{Append First} allocate from machines that already have GPUs occupied. 
It helps to reduce fragmentation. It also largely implies
that the stage is likely to be within the same machine.

\paragraph{Scatter First} try to use available GPUs equally from all used
machines, or use GPUs equally from all machines if they are all fresh.
It is suitable for those stages that have negligible weights
compared to activation sizes (less intra-stage communication).
This policy could also serve as an intermediate state to allocate GPU with minimal
fragmentation.

The overall device placement policies reduce search space effectively down
to less than $O(2^S)$, while retaining room for potential performance gain.
% \TODO{rephrase the above without O(xxx)}

%In the Formulation section \ref{sec:planner-formulation}, we will use a Pseudo
%function $D(gids, n)$ to denote the returned results of our device placement
%policies, when requested $n$ GPUs from the states $gids$.

\vspace{-.5ex}
\subsection{Planning Algorithm}
\vspace{-.5ex}
%Dynamic Programming is introduced here to find the solution with the shortest
%batch latency \(L\), with value function defined above.

Our planning algorithm use Dynamic Programming to find the optimal partition, replication and 
placement strategy, so that the pipeline latency $L$ is minimized. Here we first present 
how to update the \textit{pivot stage} ID $Q$ along the planning process, and then the 
formulation of our algorithm

\subsubsection{Determining The Pivot Stage $Q$}
%\label{sec:planner-best-q}

It is vital to select a proper pivot stage $Q$ for the estimation of $L$. 
The insight is to find the stage with minimum bubbles, which dominates steady phase.
We use a heuristic to determine $Q$ (formula \ref{eq:pivot-stage}).

\vspace{-1.5ex}
\begin{equation}
  \begin{split}
    Q & = \arg \max_{s=S-1}^{0} \max \Big( T_{st}^Q + \sum_{s'=s+1}^{Q-1} (F_{s'}+B_{s'}), T_{st}^s\Big)
  \end{split}
  \label{eq:pivot-stage}
\end{equation}

% \TODO{check they are equivalent, before removing one}

% \renewcommand{\algorithmicrequire}{ \textbf{Input:}} %Use Input in the format of Algorithm
% \renewcommand{\algorithmicensure}{ \textbf{Output:}} %UseOutput in the format of Algorithm
% \begin{algorithm}
% \caption{Iteratively update Q}
% \label{alg:iterative-q}
% \begin{algorithmic}[1]

% \STATE {let Q = S-1, s = Q-1}
% \WHILE{$s \ge 0$}
% \STATE{let $l_1$ = $(M-1) \times (F_s + B_s)$}
% \STATE{let $l_2$ = $(M-1) \times (F_Q + B_Q)$}
% \FOR{$s'=s+1$; $s' < Q$; $s'=s'+1$}
% \STATE{$l_2$ += $F_{s'} + B_{s'}$}
% \ENDFOR
% \IF{$l_1$ $>$ $l_2$}
% \STATE{$Q = s$}
% \ENDIF
% \STATE{$s = s - 1$}
% \ENDWHILE
% \end{algorithmic}
% \end{algorithm}

The initial $Q$ is set to $S-1$. \emph{DAPPLE} updates $Q$ iteratively from stage $S-1$ to stage $0$ 
according to formula \ref{eq:pivot-stage}. $T_{st}^j = (M-1) \times (F_j + B_j)$ means the duration of
steady phase, without bubbles, suppose pivot stage is $j$. For a stage $s<Q$, if
$T_{st}^s$ is larger than the sum of $T_{st}^Q$ and corresponding
forward/backward costs between $s$ and current $Q$, it means the steady phase
will have less bubbles if pivot stage is set as $s$ other than current $Q$. $Q$
will then be updated to $s$.

\subsubsection{Algorithm Formulation}

We define the estimated pipeline latency $T_{PL}(j, m, g)$ as the subproblem, 
for which we have planned the strategy for the first $j$ layers using $m$ GPUs (with device id set $g$). 
The unplanned layers forms the last stage and replicates on the other $(G-m)$ GPUs.
Our objective is to solve for $T_{PL}(N, G, \mathcal{G}), \mathcal{G}=\{0, 1, ..., G-1\}$.
$N$, $G$ and $\mathcal{G}$ denote the number of layers, number of GPUs and GPU
set, respectively. Formula~\ref{eq:planning_algorithm} describes the algorithm.

% Our DP algorithm tries to split the layers into two parts,
% with the second part being a single stage and recursively partition the first
% part. For each split, the algorithm enumerates the number of GPUs allocated to
% the last stage, and use the above three strategy for device placement. We have:
% The
% current best split, replication and placement is recorded along the planning
% process. \TODO{rewrite it.}

% The dynamic programming tries to partition the model from the first layer to the whole model with 
% placement and replication attemption iteratively. We have:

\vspace{-1.5ex}
\begin{equation}
  \left.
      T_{PL}(N, G, \mathcal{G})
  \right.
  = \min_{1 \le j < N} \min_{1 \le m < G} \min_{g \in D(\mathcal{G}, m)}
  \left.
      T_{PL}(j,m,g)
  \right.
  \label{eq:planning_algorithm}
\end{equation}

\begin{figure}[!tbp]
  \centering
  \includegraphics[width=.9\linewidth]{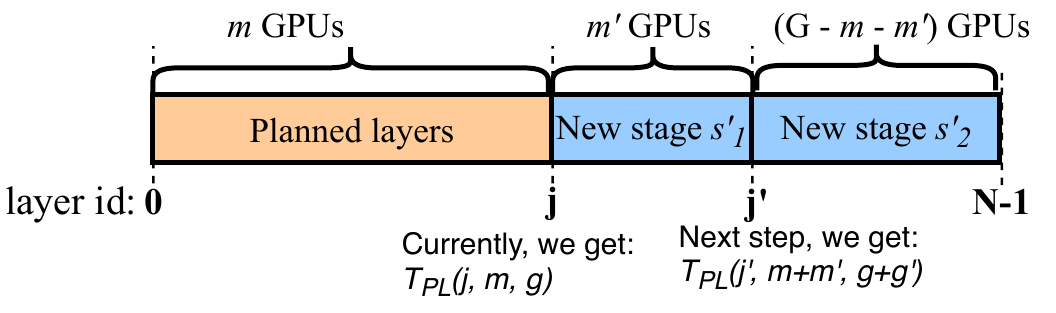}
  \caption{\label{fig:HPGO-Planning-Process-jprime}
    Planning process for j'.}
\end{figure}

Fig. \ref{fig:HPGO-Planning-Process-jprime} describes the iterative planning process.
Suppose we have already planned for the first $j$ ($0 \leq j < N$) layers and have the 
estimation $T_{PL}(j, m, g)$ as pipeline latency. The layers after $j$ forms a stage $s'$.
Meanwhile, we get the optimal $Q$ for current strategy along with the 
cost of $F_{Q}$ and $B_{Q}$ for stage $Q$.
Next step, we try to add one more partition in stage $s'$, supposing after 
layer $j'$ ($j < j' \leq N$), and split $s'$ into two new stages $s'_1$ and $s'_2$. 
We assign $m'$ GPUs for $s'_1$ and $(G-m-m')$ GPUs for $s'_2$, and estimate $T_{PL}(j', m+m', g+g')$ 
according to formula \ref{eq:one_step_iter}. Note \emph{DAPPLE} enumerates the three strategies 
in section \ref{subsec:device_assignment} for device placement of stage $s'_1$.

\vspace{-1.5ex}
\begin{equation}
\begin{split}
  T_{PL}(j', m+m', g+g') = L &
\end{split}
\label{eq:one_step_iter}
\end{equation}

Here, $L$ is the same with that in formula \ref{eq:latency-objective}. 
The key for the estimation of $L$ in formula \ref{eq:one_step_iter} is to find $Q$ of 
subproblem $T_{PL}(j', m+m', g+g')$. In the sample in Fig. \ref{fig:HPGO-Planning-Process-jprime}, 
we get $Q_j$ for $T_{PL}(j, m, g)$. We apply formula \ref{eq:pivot-stage} to 
get $Q_{j'}$ for $T_{PL}(j, m+m', g+g')$ with the help of $Q_j$:
if $Q_j$ is not $s'$, we do not need to iterate all stages before $j$, 
but use $Q_j$ for all stages before layer $j$ instead in the iterative process.

Along the above process, we record the current best split, replication and
placement for each point in our solution space using memorized search.

\vspace{-.5ex}
% {\color{blue}
\subsection{Contributions over previous work}
\label{sec:contribution-over-previous-work}
\vspace{-.5ex}

Previous works on pipeline planning includes PipeDream\cite{narayanan2019pipedream} 
(for asynchronous training) and torchgpipe\cite{torchgpipe}, a community implementation of
GPipe\cite{huang2019gpipe} which uses ``Block Partitioning of Sequences''
\cite{barany2015block}. Both aim to balance the workload across all
GPUs. While this idea works good in PipeDream's asynchronous
scenarios and gives reasonable solutions under GPipe's synchronous pipeline for 
its micro-batch arrangement, we found that our micro-batch arrangement could
achieve even higher performance by 1) intentionally preferring slightly uneven
partitioning with fewer stages, and 2) exploring a broader solution 
space of device assignment. The following sections highlight our contributions 
of planning for hybrid parallelism. The resulting strategies and performance
gain on real-world models will be demonstrated in Section \ref{sec:comparison-with-pipedream}.

\subsubsection{Uneven Pipeline Partitioning with Fewer Stages}

\begin{figure}[!tbp]
  \centering
  \includegraphics[width=0.9\linewidth]{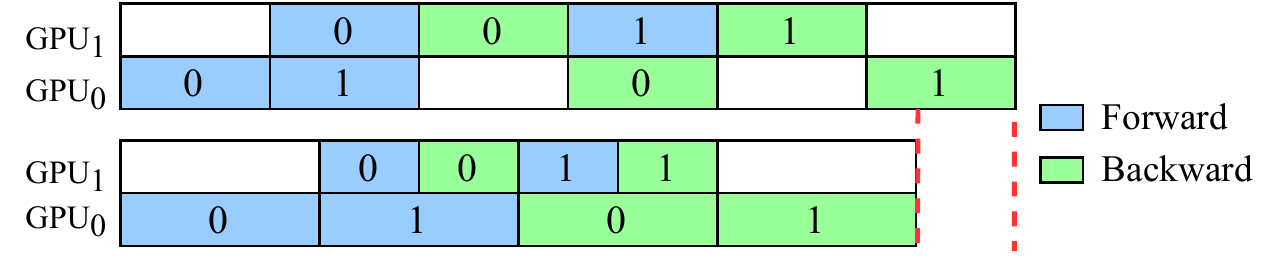}
  \caption{\label{fig:Uneven-Pipeline-MWE}
    Uneven pipeline minimum example.}
\end{figure}

In synchronous Pipeline Parallelism scenarios, we found two insights that
could provide an additional performance improvements. The first one is to
partition the model into as few stages as possible to minimize the bubble
overhead under the same number of micro-batches. This conclusion is also
mentioned in GPipe. The second one is that partitioning
the model in a slightly uneven way yields much higher performance than a
perfectly even split, like the example in Fig. \ref{fig:Uneven-Pipeline-MWE}.

\subsubsection{Versatile Device Placement}

\emph{DAPPLE} device assignment strategy covers a broad solution space 
for stage placement, and is a strict superset of PipeDream's
hierarchical recursive partitioning approach. This allows us to 
handle various real world models. 
For example, for models that have layers with huge activations compared to
their weights, \emph{DAPPLE} allows such a layer to be replicated across
multiple machines (Scatter First) to utilize high-speed NVLink for activation 
communication and low-speed Ethernet for AllReduce.

% }
\section{\emph{DAPPLE} Runtime}
\label{sec:impl}

\subsection{Overview}

We design and implement \emph{DAPPLE} runtime in Tensorflow\cite{tensorflow} (TF) 1.12,
which employs a graph based execution paradigm. As common practices, TF takes 
a precise and complete computation graph (DAG), schedules and executes
graph nodes respecting all data/control dependencies.

\emph{DAPPLE} runtime takes a user model and its planning results as input,
transforms the model graph into a pipelined parallel graph and executes on multiple
distributed devices. 
%Fig. \ref{fig:impl-overview} shows key components of the
%transformed pipelined graph for a single micro batch graph unit.
%\begin{figure}
%    \includegraphics[width=0.45\textwidth]{./implementation/figs/micro-batch-graph.png}
%    \caption{
%    Per Micro Batch Pipeline Graph
%    }
%    \label{fig:impl-overview}
%\end{figure}
%The planner generates a detailed partitioning of pipeline stages, and the mapping from
%each stage to a target set of devices. 
%The runtime transforms the input model graph 
It first builds forward/backward graphs separately for each pipe stage. 
Then additional split/concat nodes are introduced between adjacent stages for
activation communication. Finally, it builds a subgraph to perform weights update for
synchronous training. This step is replication aware, meaning it will generate 
different graphs with or without device replication of stages.
We leverage control dependences in TF to enforce extra execution orders among forward/backward stage computations.
%which is a key component to enable micro-batch scheduling. 
%In this work, we implement this pipelined graph construction process by manually manipulating computation nodes
%of user models. This graph transformation can be done through an automatic tool, which is left as our future work.
Section \ref{sec:building-micro-batch-unit} presents how to build basic TF graph units for a single micro-batch.
Section \ref{section:micro-batch-schedule} discusses how to chain multiple such units using control dependencies
to facilitate \emph{DAPPLE} execution.

\subsection{Building Micro-batch Units}
\label{sec:building-micro-batch-unit}

\subsubsection{Forward/Backward Stages}

%\begin{algorithm}
%    \caption{Building Micro Batch Unit}
%    \label{alg:micro_batch_unit}
%    \begin{algorithmic}[1]
%    \REQUIRE ~~ \\
%    $I$: One micro batch data \\
%    $S = [(l_i, d_i), (l_{i+1}, d_{i+1}), ...]$: Tuples of layers in each stage with their assigned device \\
%    \ENSURE ~~ \\
%    $A$: Array of activations after each stage \\
%    $G_v$: Array of gradients about trainable variables
%    \STATE{$I_p = I, N=length(S), g_p = None$}
%    \FOR{$(l, d)$ in $S$}
%    \STATE{Switch to device $d$}
%    \STATE{$out = l$.forward($I_p$)}
%    \STATE{Append $out$ to $A$}
%    \STATE{$I_p = out$}
%    \ENDFOR
%    \FOR{$i=N-1$; $i>0$; $i--$}
%    \STATE{$l, d = S[i]$, $a = A[i]$, $a_p=A[i-1]$}
%    \STATE{Switch to device $d$}
%    \STATE{Collect trainable variables $v_i$ in stage $i$}
%    \STATE{$g_a, g_v$ = tf.gradients($a, [a_p, v_i], g_p$)}
%    \STATE{Append $g_v$ to $G_v$}
%    \STATE{$g_p = g_a$}
%    \ENDFOR
%    \STATE{Switch to device $d$ refers to $A[0]$}
%    \STATE{Collect trainable variables $v$ refers to $A[0]$}
%    \STATE{$g$ = tf.gradients($A[0], v, g_p$)}
%    \STATE{Append $g$ to $G_v$}
%    \end{algorithmic}
%\end{algorithm}

In order to enforce execution orders with control dependencies between stages,
we need to build forward and backward graphs stage by stage to deal with the 
boundary output tensors such as activations.

Specifically, we first construct the forward graph of each stage in sequence and record the boundary tensors.
No backward graphs should be built until all forward graphs are ready. Second, backward graphs will be 
built in reverse order for each stage.

%\emph{tf.gradients} API
%should be called when generating backward graphs in reverse order on each stage instead of 
%\emph{optimizer.compute\_gradients} \TODO{what is the purpose here? Do not list TF API. No reviewer knows
%what the API means.. Just explain the purpose of the API instead.}. 
%Algorithm \ref{alg:micro_batch_unit} decribe this procedure.
%We refer interested readers to our github for such graph manipulation details.
% we first remove all \emph{optimizer.compute\_gradients} calls defined in the input 
% model. These are backward specific construction APIs. We won't start building 
% backward graphs explicitly unless all forward graphs are ready.
% We build forward graphs stage by stage. After building computation nodes for each
% stage, we record all weights and activation/loss output tensors. 
% Weights information is necessary to build nodes for pipelined weight updates.
% Activation/Loss tensors are necessary to backward graph generation.
% In micro batch scheduling, such boundary tensors are essential to enforce execution
% orders with control dependencies between computations of different stages.

% Once all graphs for forward stages are available, we start building graphs for
% backward stages in reverse order. We follow TF auto differentiation rules.
% For each backward stage, we explicitly call \emph{tf.gradients}. It takes output
% tensors of the corresponding forward stage as inputs, and builds a backward graph
% for the stage.
% We refer interested readers to our Github for such graph manipulation details.

\subsubsection{Cross Stage Communication}

\begin{figure}[!tbp]
  \centering
    \includegraphics[width=\linewidth]{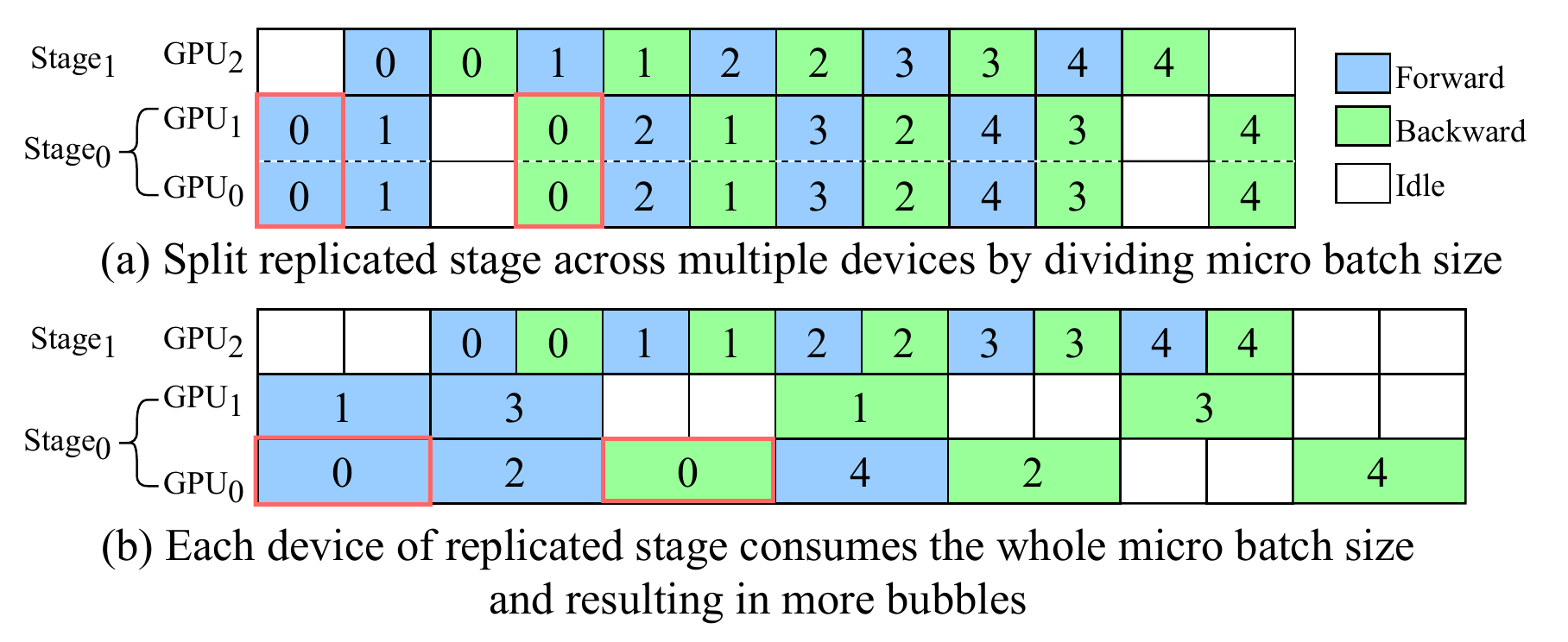}
    \caption{
    Efficiency of two stage replication approaches. 
    Stage 0 consumes twice as much time as stage 1 for a micro-batch.
    }
    \label{fig:split-concat-impl}
\end{figure}

\emph{DAPPLE} replicates some stages such that the number of nodes running a stage
can be different between adjacent stages, and the communication patterns between them are different from straight pipeline design.
We introduce special \emph{split-concat} operations between these stages.

Fig. \ref{fig:split-concat-impl}(a) shows the replication in \emph{DAPPLE} for a 
2-stage pipeline, whose first stage consumes twice as much time as the second stage for a micro-batch
and thus is replicated on two devices.
For the first stage, we split the micro-batch further into 2 even slices, and
assign each to a device.
An alternative approach\cite{narayanan2019pipedream} (Fig. \ref{fig:split-concat-impl}(b))
is not to split, but to schedule an entire micro-batch to two devices in 
round robin manner. However, the second approach has lower pipeline efficiency due to tail effect\cite{taileffect}.
Though the second approach does not involve extra split-concat operations, the overhead of
tail effect is larger than split-concat in practice.
We hence use the first approach with large enough micro-batch size setting to ensure device efficiency.

%Both methods require params reduction across replicated devices at the up- date phase, while the latter introduces more bubbles when same micro_batches are injected. DAPPLE takes scheduling (a) for its better system efficiency and ease-to-implement on Tensorflow. Note that extra split and concat nodes need to be inserted to the graph of DAPPLE unit, as shown in the top right of Figure 14 while the cost is negligible. The placement of split op is co-placed with its inputs while the concat op is co-placed with its outputs for performance reasons. For example, in VGG19 15:1 straight pipeline’s graph generation, the concat op placed on the first stage can be 80% slower than be placed with the second stage for extra communication of activations (as well as gradients) are introduced.

\begin{figure}
  \centering
    \includegraphics[width=\linewidth]{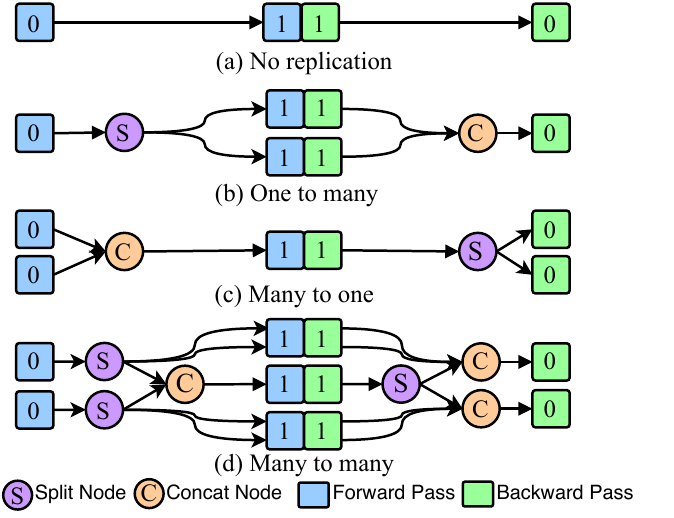}
    \caption{
    %Communication primitives at partition boundary for replicated stage(s)
    Split-Concat for Cross Stage Communication
    }
    \label{fig:replicate-approach}
\end{figure}

The split-concat operations include \textit{one-to-many}, \textit{many-to-one} and \textit{many-to-many} communication. We need \texttt{split} for \textit{one-to-many}(Fig. \ref{fig:replicate-approach}(b)), that is, splitting the micro-batch into even slices and sending each slice to a device in the next stage. We need \texttt{concat} for \textit{many-to-one}(Fig. \ref{fig:replicate-approach}(c)), where all slices should be concatenated from the previous stage and fed into the device in the next stage. For \textit{many-to-many}(Fig. \ref{fig:replicate-approach}(d)) we need both \texttt{split} and \texttt{concat} of micro-batch slices to connect adjacent stages. If two adjacent stages have the same replication count, no \emph{split-concat} is needed.

\subsubsection{Synchronous Weights Update}

\begin{figure}
  \includegraphics[width=0.45\textwidth]{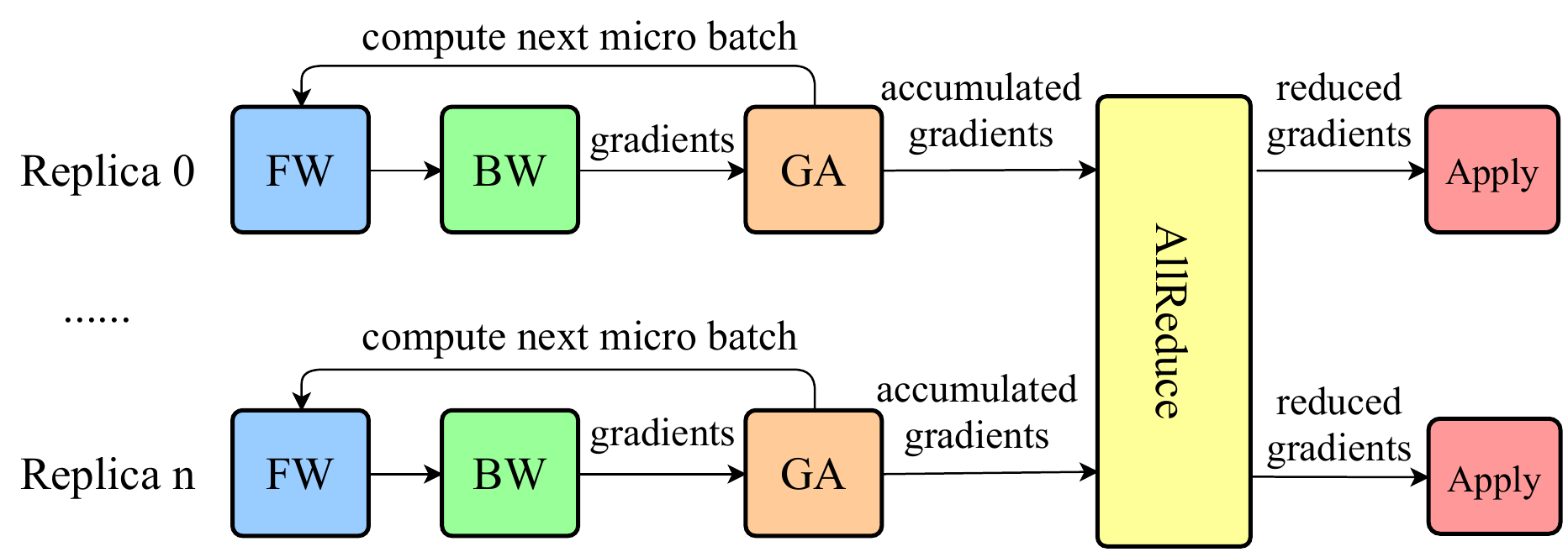}
  \caption{
  Weights update. GA means gradient accumulation\cite{GA-TF}.
  }
  \label{fig:weight-updates}
\end{figure}

Weights updating in \emph{DAPPLE} is different with naive training as there are multiple micro-batches 
injected concurrently. Meanwhile, the replication makes weights updating more complex.
As is shown in Fig. \ref{fig:weight-updates}, each device produces and accumulates gradients for all micro-batches.
There is an \emph{AllReduce} operation to synchronize gradients among all replicas, if exists.
A normal \emph{Apply} operation updates weights with averaged gradients eventually.

\subsection{Micro-batch Unit Scheduling}
\label{section:micro-batch-schedule}

%\emph{DAPPLE} constructs a single computation graph with all micro batches interleaved with each other. 
The \textit{early backward scheduling} strikes a trade-off between 
micro-batch level parallelism and peak memory consumption: feeding more micro-batches into pipeline at once
implies higher parallelism, but may lead to more memory usage.

\emph{DAPPLE} scheduler enforces special execution orders between micro-batches to reduce 
memory usage. For the first stage, we suppose $K$ micro-batches are scheduled concurrently at the beginning 
for forward computation. Specifically, $K_i$ is the number of scheduled micro-batches at the beginning for stage $i$.
The overall execution follows a round robin order with interleaving FW and BW.

\begin{figure}
  \centering
    \includegraphics[width=\linewidth]{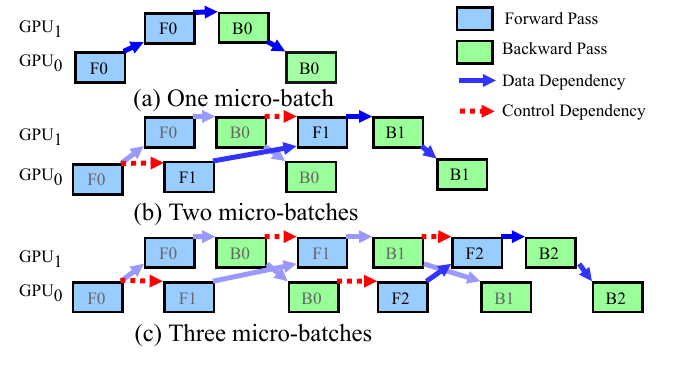}
    \caption{
    Micro-batches scheduling. The solid blue and dotted red arrows denote data and control 
    dependencies, respectively.
    }
    \label{fig:micro-batch-schedule}
\end{figure}

We realize the scheduler with control dependency edges in TF. Fig. \ref{fig:micro-batch-schedule} 
shows how up to three micro-batches are connected via control dependencies to implement the schedule 
for a two stage pipeline.
Control dependency is not necessary when there is only one micro-batch (Fig. \ref{fig:micro-batch-schedule}(a)). 
With two micro-batches (Fig. \ref{fig:micro-batch-schedule}(b)), two control edges are introduced.
The control edge between $B0$ and $F1$ in stage $1$ is to form the round robin order of FW 
and BW. The early execution of $B0$ helps to free memory of $F0$ and $B0$ in stage 1, 
which can be reused in following tasks. The edge between $F0$ and $F1$ in stage $0$ 
is to enforce the order that micro-batch 1 is strictly executed after micro-batch 0, thus the 
backward of micro-batch 0 can be executed earlier and its corresponding memory can be freed earlier. 
In practice, $F0$ and $F1$ are typically large chunks of computations. The lack of parallelism 
between $F0$ and $F1$ does not affect the gain of reducing memory usage. The situation with three micro-batches 
(Fig. \ref{fig:micro-batch-schedule}(c)) is the same.

% Following are rules for inserting control edges for stage $i$:
% \begin{itemize}
% \item[-] In the warm up phase, for each forward micro-batch $F_k$ ($0 < k < K_i$), 
% there is a control edge between $F_{k-1}$ and $F_k$.
% \item[-] After warm up phase, for each backward micro-batch $B_m$ ($0 \le m < M-1$),
% there is a control edge between $B_m$ and $F_{m+S-i}$. $M$ and $S$ denote the number of 
% micro-batches scheduled in a training iteration, and the number of stages, respectively.
% \end{itemize}

An appropriate $K_i$ is essential as it indicates the peak memory consumption for stage $i$.
There are two primary factors for deciding $K_i$: memory demand for one micro-batch execution, and 
the ratio between cross stage communication latency and the average
FW/BW computation time (referred as {\em activation communication ratio, ACR in short}). 
The former determines how many forward batches can be scheduled concurrently at most. 
We define the maximum number of micro-batches supported by the device memory as $D$;
Lower ACR means less warm up forward batches $K_i$ (smaller $K_i$) are enough to saturate the pipeline. 
While notable ACR means larger $K_i$ is necessary.

We implement two policies to set $K_i$ in practice.
\textit{Policy A} ($P_A$): $K_i = min(S-i, D)$. $P_A$ works well when 
ACR is small, i.e. the impact of cross stage communication overhead is negligible.
\textit{Policy B} ($P_B$): $K_i = min(2*(S-i)-1, D)$. Here we schedule twice the
number of forward micro-batches than $P_A$. The underlying intuition is
that in some workloads, the cross stage communication overhead is comparable with
forward/backward computations and thus more micro-batches is needed to saturate the pipeline.

% In our experiments, we do observe two workloads (VGG-19 and AmoebaNet-36 in section \ref{sec:scheduling-policy}),
% for which $P_B$ works better.
% In both polices, we keep $K_i$ at $O(S)$. 
% In practice, the number of pipe stages ($S$) produced by the \emph{DAPPLE} planner is typically much smaller 
% than the number of micro-batches ($M$). This is important to constrain peak memory consumption.
% A further implication is that \emph{DAPPLE} encourages relatively large granularity of stage computations, 
% thus achieving decent execution efficiency.
%This is important because short pipelines work better in most of our cases 
%\TODO{Why?}\TODO{DO we really need to elaborate this further?}.

\section{Evaluation}
\label{section:evaluation}

%\TODO{One problem to address in introduction: how to chose the parallelism scheme between DP and PP.}

\subsection{Experimental Setup}
\label{sec:exp-setup}

% \vspace{1mm}
\noindent\textbf{Benchmarks.} Table \ref{tbl:models} summarizes all the six representative DNN models that we use as benchmarks
in this section.
The datasets applied for the three tasks are WMT16 En-De \cite{sennrich2016edinburgh},
SQuAD2.0 \cite{rajpurkar2018know} and ImageNet\cite{russakovsky2015imagenet}, respectively.
\begin{table}[t]
    \caption{
     Benchmark models. %$PBS$ is short for profile batch size.
    }
    \label{tbl:models}
    \begin{center}
    \begin{tabular}{llcl}
    \toprule
    Task & Model & \makecell[c]{\# of\\ Params} & \makecell[c]{(\emph{cbch Size},\\ Memory Cost)} \\ %Dataset \\%& \makecell[c]{max batch\_size\\on V100}
    \midrule
    Translation & GNMT-16 \cite{wu2016google} & 291M & (64, 3.9GB) \\ %\makecell{WMT16\\En-De} \\
    % \multirow{2}{*}{Translation} & GNMT-16 & 291M &\multirow{2}{*}{\makecell{WMT16\\EN-De}} \\
    % & GNMT-32 & 459M & \\
    \cline{1-4}
    \multirow{2}{*}{\makecell[l]{Language\\Model}} & BERT-48 \cite{devlin2018bert} & 640M & (2, 11.4GB)\\ %\multirow{2}{*}{SQuAD2.0 \cite{rajpurkar2018know}}\\
    & XLNet-36 \cite{yang2019xlnet}& 500M & (1, 12GB) \\
    \cline{1-4}
    \multirow{3}{*}{\makecell[l]{Image \\Classification}} & ResNet-50 \cite{he2016deep} & 24.5M & (128, 1GB)\\ %\multirow{4}{*}{ImageNet\cite{russakovsky2015imagenet} } \\
    & VGG-19 \cite{simonyan2014very}& 137M & (32, 5.6GB)\\
    % & AmoebaNet-6 & 176M &\\
    % & AmoebaNet-18 & 479M &\\
    & AmoebaNet-36 \cite{shah2018amoebanet}& 933M & (1, 20GB)\\
    \bottomrule
    \end{tabular}
    \end{center}
\end{table}
% Both AmoebaNet variants have filter sizes of 512, but with different (18 VS. 36)
% configurations for normal cell layers.
% Table \ref{tbl:models} summarizes this.

\begin{table}[t]
    \caption{Hardware configurations.}
    \label{tbl:cluster-spec}
    \begin{center}
    \begin{tabular}{cccc}
    \toprule
    Config & \makecell[c]{GPU(s) per\\server($N_s$)} & \makecell[c]{Intra-server\\connnections} &\makecell[c]{Inter-server\\connections} \\
    \midrule
    % A & 8x V100 & NVLink & N/A \\
    A & 8x V100 & NVLink & 25 Gbps \\
    B & 1x V100 & N/A & 25 Gbps \\
    C & 1x V100 & N/A & 10 Gbps \\
    \bottomrule
    \end{tabular}
    \end{center}
\end{table}

\vspace{1mm}
\noindent\textbf{Hardware Configurations.}
Table \ref{tbl:cluster-spec} summarizes three common hardware environments
for DNN training in our experiments, where hierarchical and flat interconnections are both covered.
In general, hierarchical interconnection is popular in industry GPU data centers.
We also consider flat Ethernet networks interconnections because NVLink may 
not be available and GPU resources are highly fragmented in some real-world production clusters.
Specifically,
\emph{Config-A} (hierarchical) has servers each with 8 V100 interconnected with 
NVLink, and a 25Gbps Ethernet interface.
\emph{Config-B} (flat) has servers each with only one V100 (no NVLink) and a 25Gbps Ethernet interface.
\emph{Config-C} (flat) is the same with \emph{Config-B} except with only 10 Gbps Ethernet equipped.
The V100 GPU has 16 GB of device memory.
All servers run 64-bits CentOS 7.2 with CUDA 9.0, cuDNN v7.3 and NCCL 2.4.2\cite{nccl2019}.

\vspace{1mm}
\noindent\textbf{Batch Size and Training Setup}. 
The batch sizes of offline profiling for the benchmarks are shown in the last column 
of Table \ref{tbl:models} (\emph{ch size}.
As for AmoebaNet-36, it reaches OOM even if $batch\_size=1$ on a single V100. Thus we extend to two V100s where $batch\_size=1$ just works. % to Problem-2
We use large enough global batch size for each benchmark to ensure high utilization on each device.
All global batch sizes we use are consistent with common practices of the ML community.
We train GNMT-16, BERT-48 and XLNet-36 using the Adam optimizer
\cite{kingma2014adam} with initial learning rate of 0.0001, 0.0001, 0.01 and
0.00003 respectively.
For VGG19, we use SGD with an initial learning rate of 0.1.
For AmoebaNet, we use RMSProp \cite{rmsprop12} optimizer with an initial learning rate of 2.56.
We use \emph{fp32} for training in all our experiments.
Note that all the pipeline latency optimizations proposed in this paper give equivalent gradients for training when keeping global batch size fixed and thus convergence is safely preserved. % to Problem-10

\subsection{Planning Results}
\label{sec:dpl-results-overview}

% \begin{figure}
%     \includegraphics[width=0.47\textwidth]{./experiments/figs/scheduling_policies.pdf}
%     \caption{Normalized training throughput vs. two scheduling policies}
%     \label{fig:scheduling_policies}
% \end{figure}
\begin{table}[t]
    \caption{Normalized training throughput speedup of scheduling policies $P_B$ compared to $P_A$.}
    \label{tbl:shcedule-policy}
    \begin{center}
    \begin{tabular}{ccccc}
    \toprule
    Model & Bert-48 & XLNet-36 & VGG-19 & GNMT-16 \\
    \midrule
    Speedup & 1.0 & 1.02& 1.1 & 1.31 \\
    \bottomrule
    \end{tabular}
    \end{center}
\end{table}
\begin{figure*}[!htbp]
    \centering
    \subfloat[VGG19 on config $A$]{\includegraphics[width=0.33\textwidth]{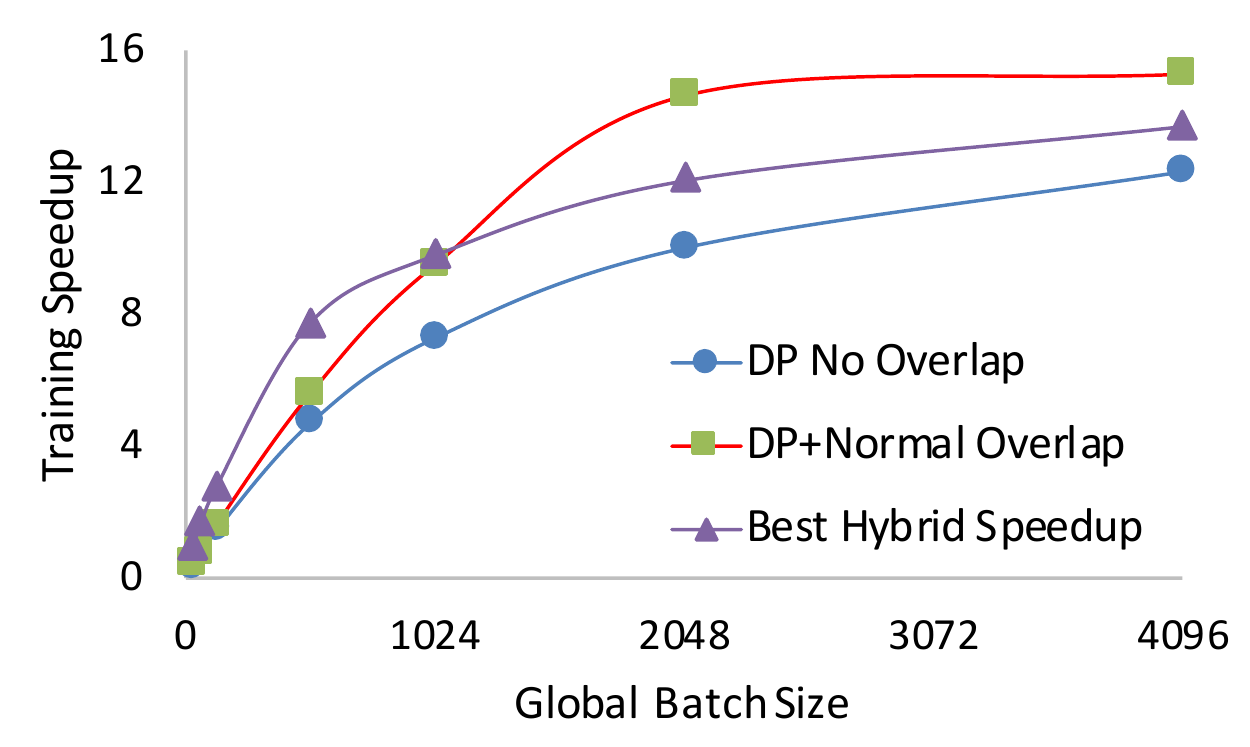}}
    \subfloat[VGG19 on config $B$]{\includegraphics[width=0.33\textwidth]{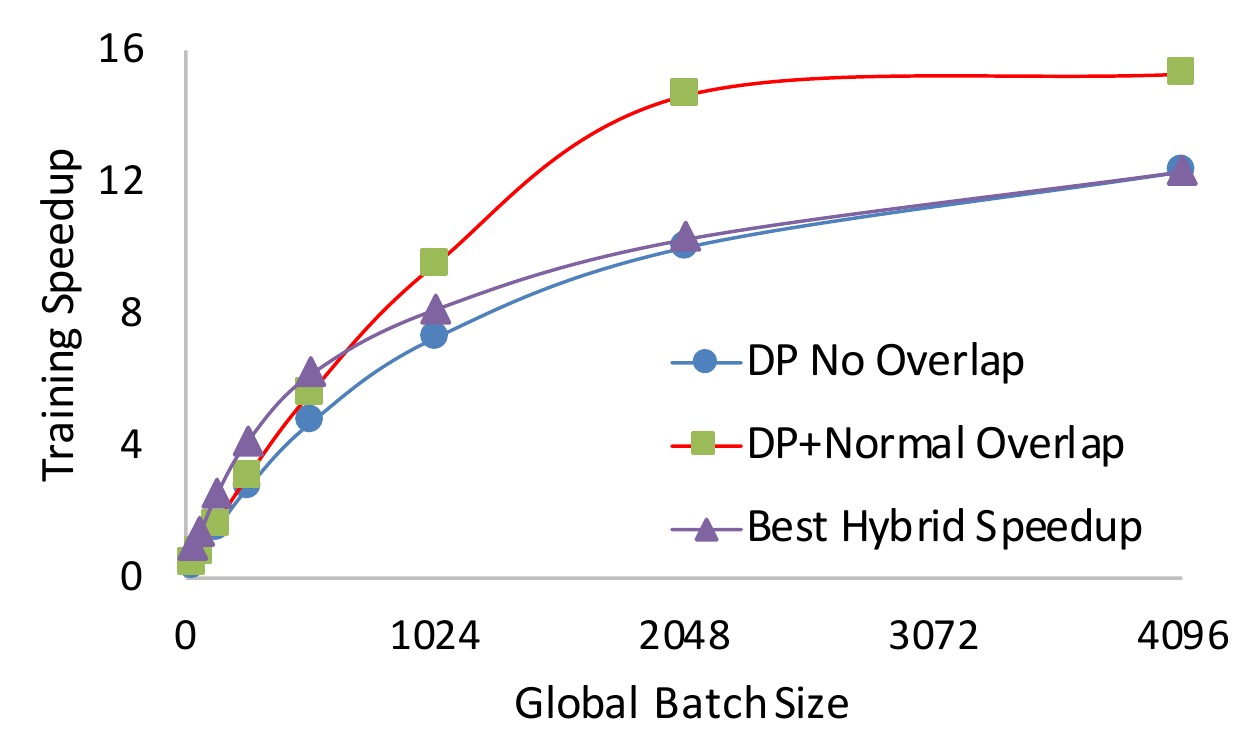}}
    \subfloat[VGG19 on config $C$]{\includegraphics[width=0.33\textwidth]{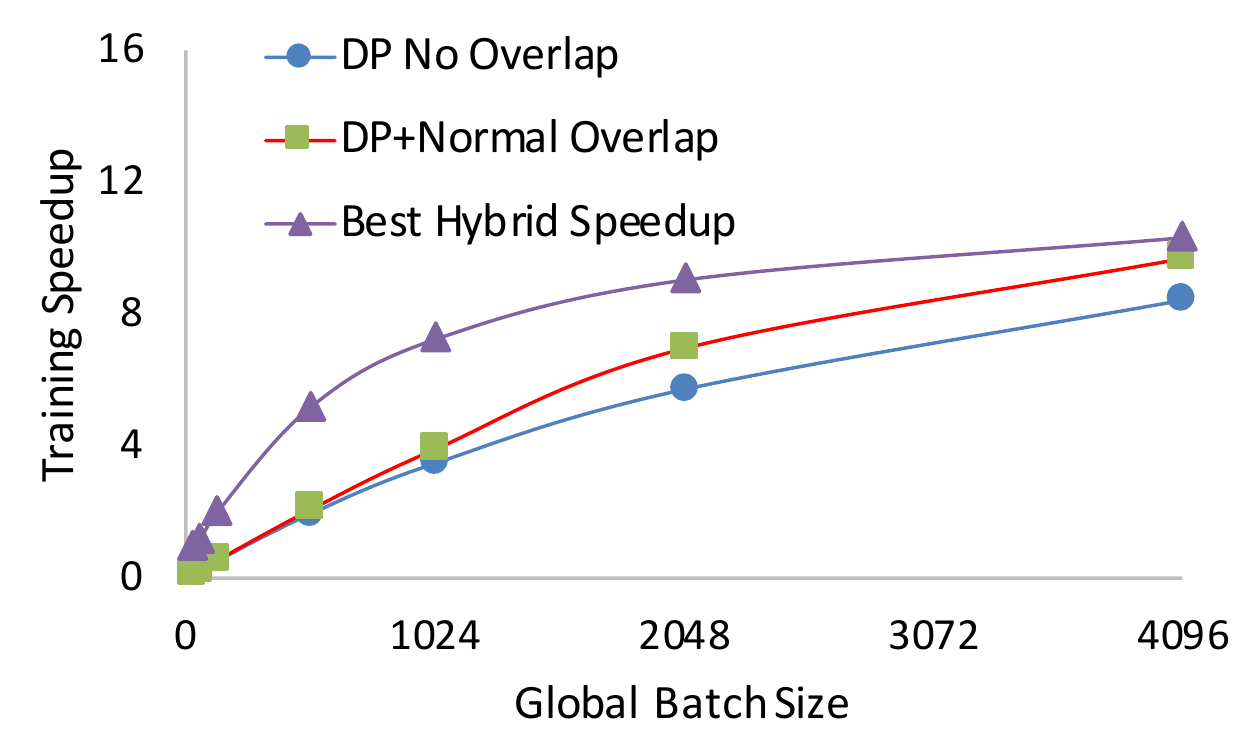}} \\
    \subfloat[GNMT-16 on config $A$]{\includegraphics[width=0.33\textwidth]{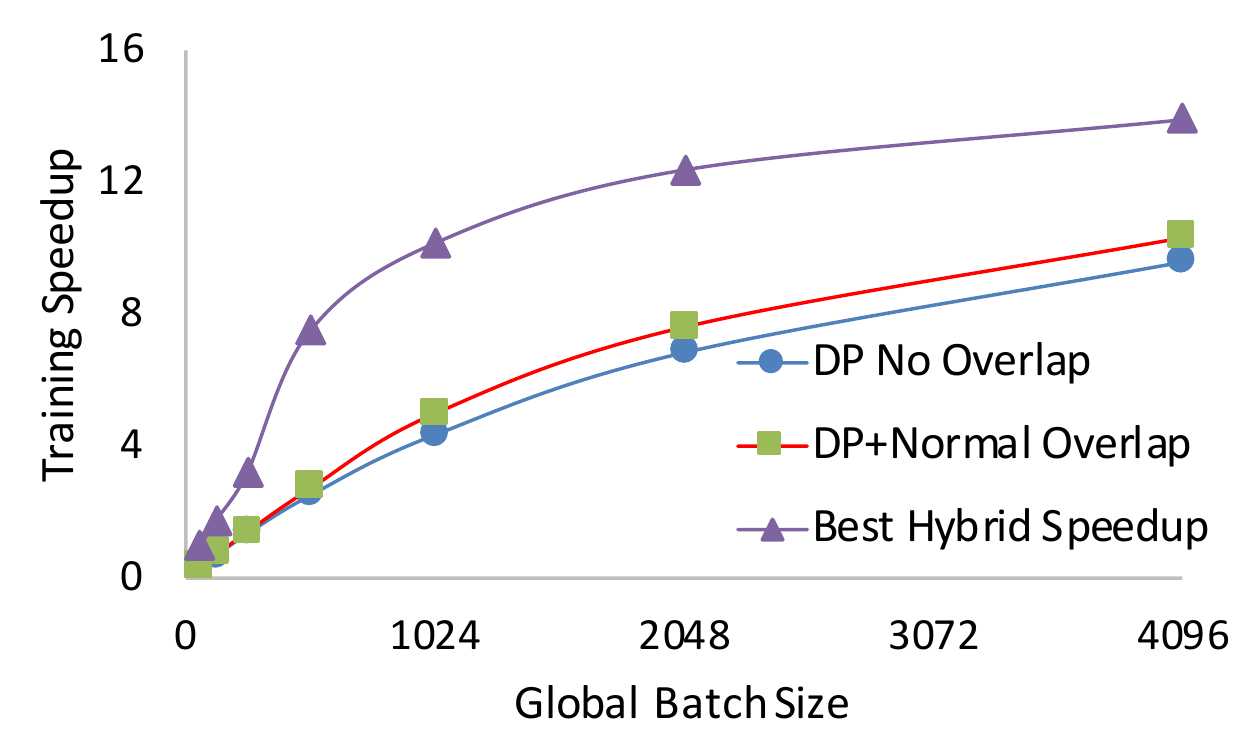}}
    \subfloat[GNMT-16 on config $B$]{\includegraphics[width=0.33\textwidth]{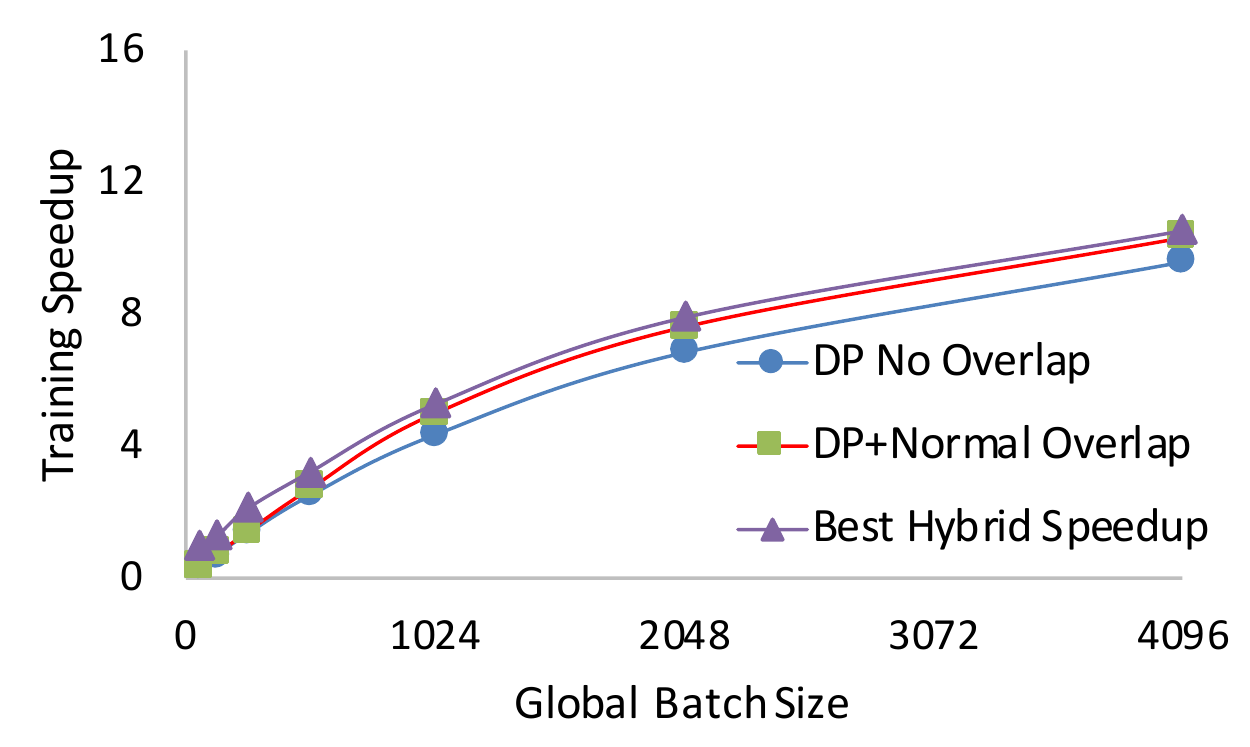}}
    \subfloat[GNMT-16 on config $C$]{\includegraphics[width=0.33\textwidth]{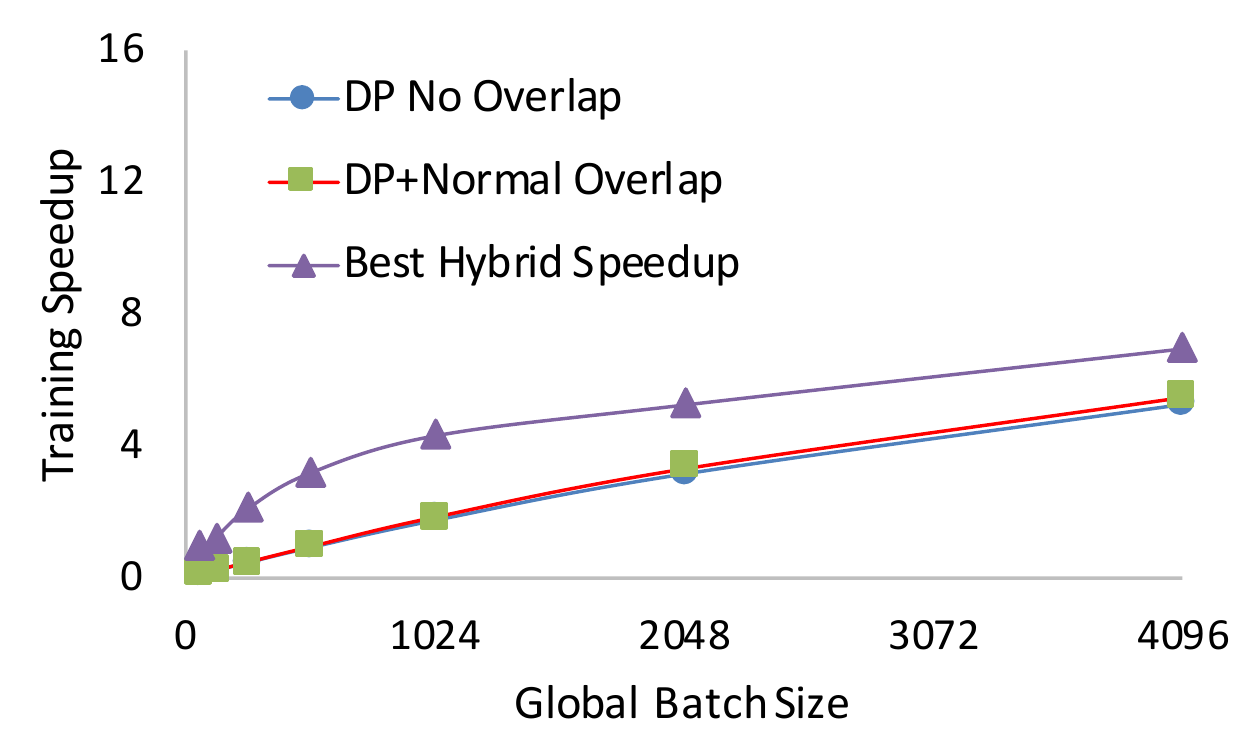}} \\
    \subfloat[BERT-48 on config $A$]{\includegraphics[width=0.33\textwidth]{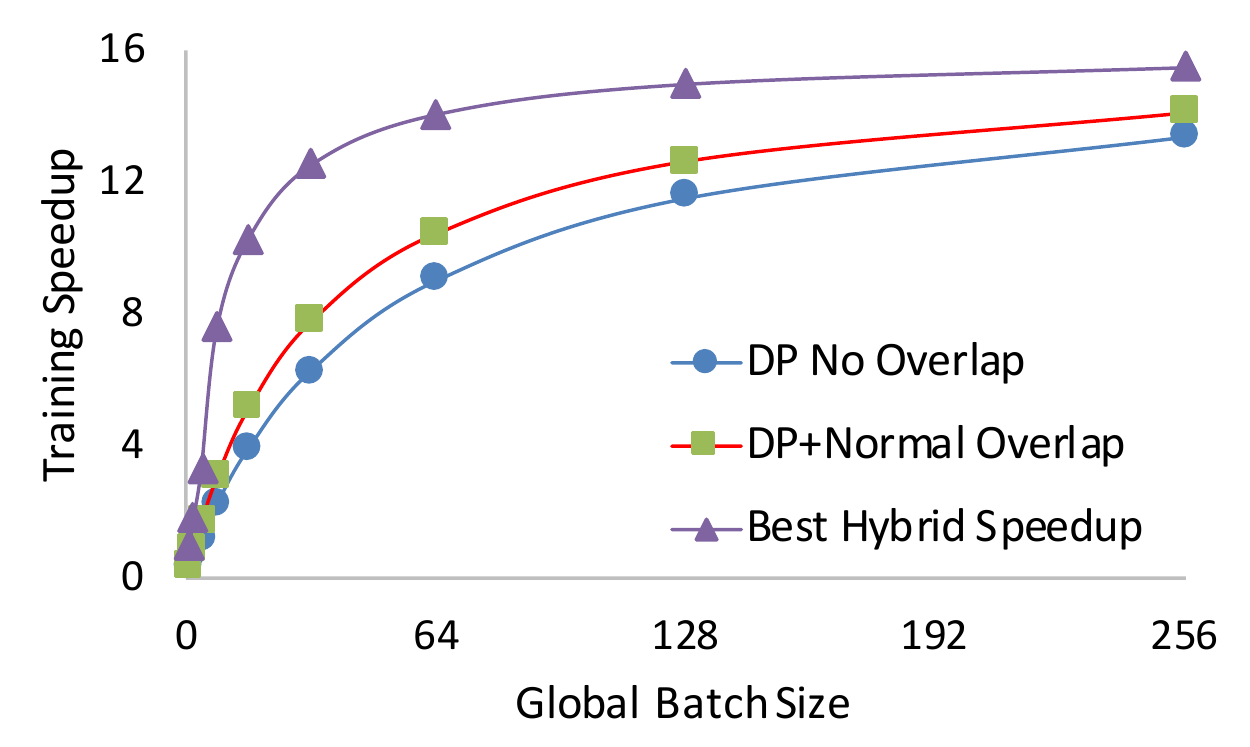}}
    \subfloat[BERT-48 on config $B$]{\includegraphics[width=0.33\textwidth]{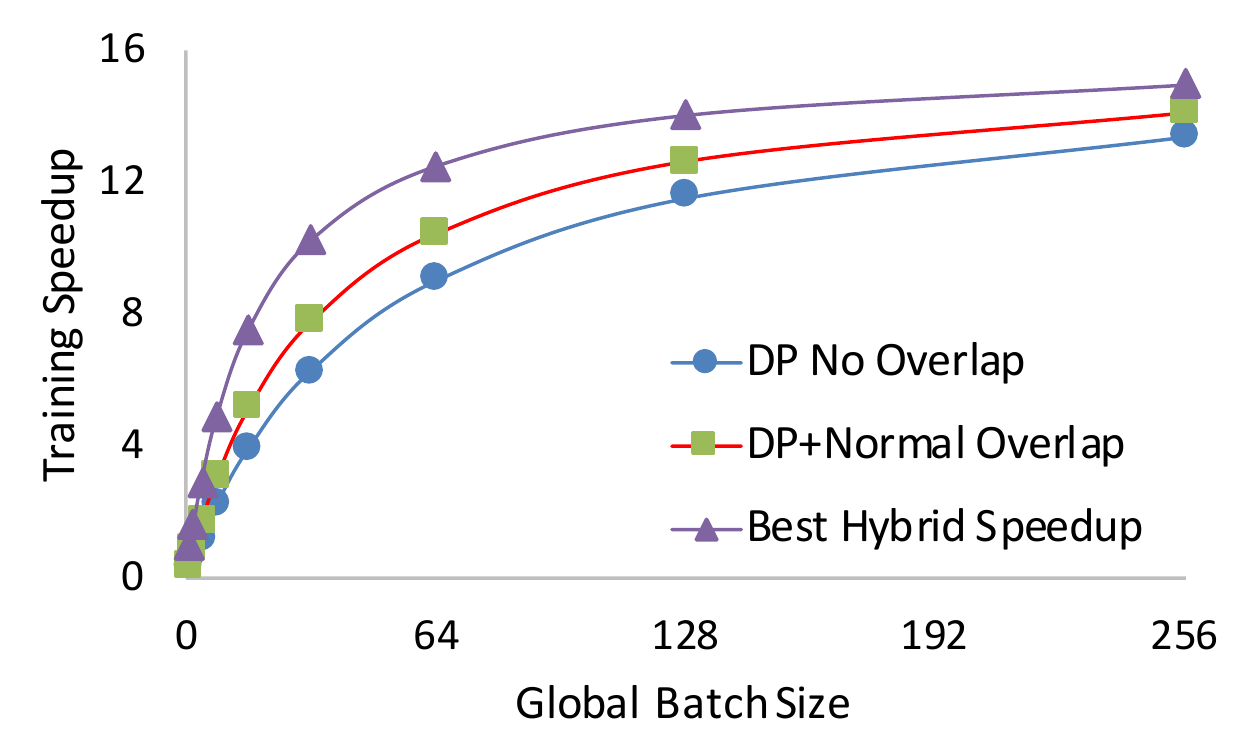}}
    \subfloat[BERT-48 on config $C$]{\includegraphics[width=0.33\textwidth]{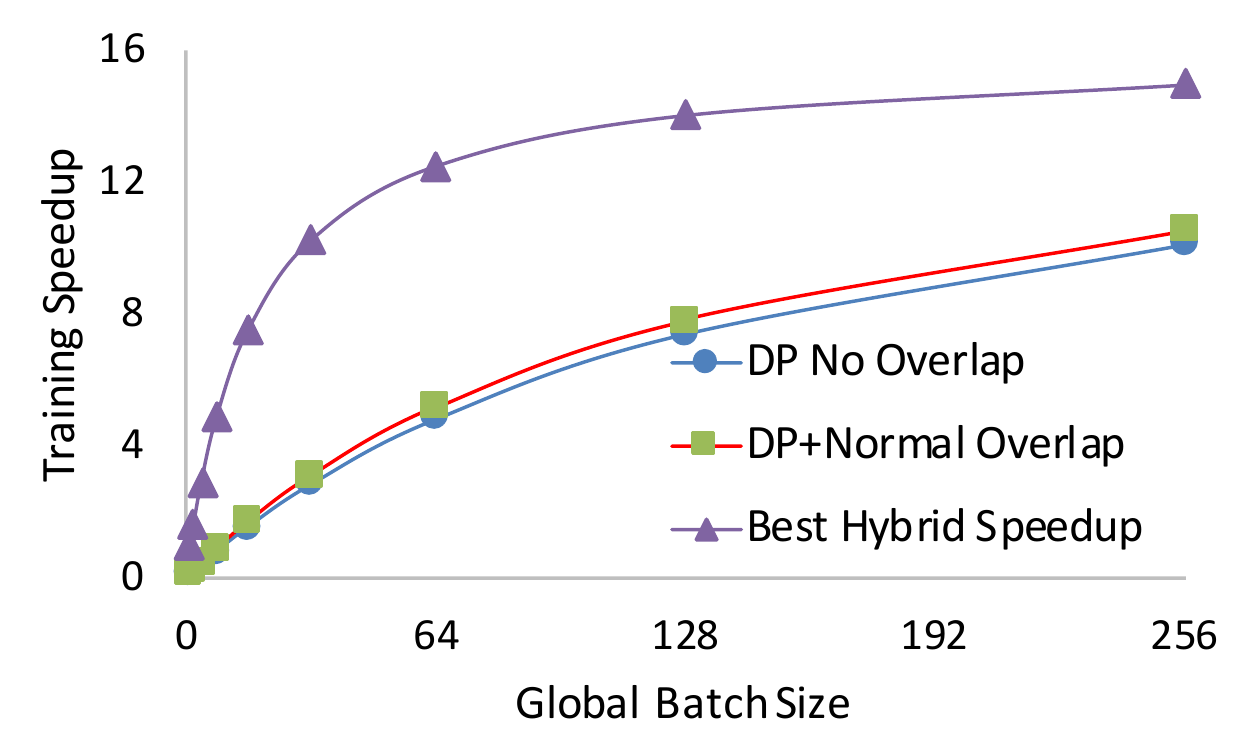}} \\
    \subfloat[XLNet-36 on config $A$]{\includegraphics[width=0.33\textwidth]{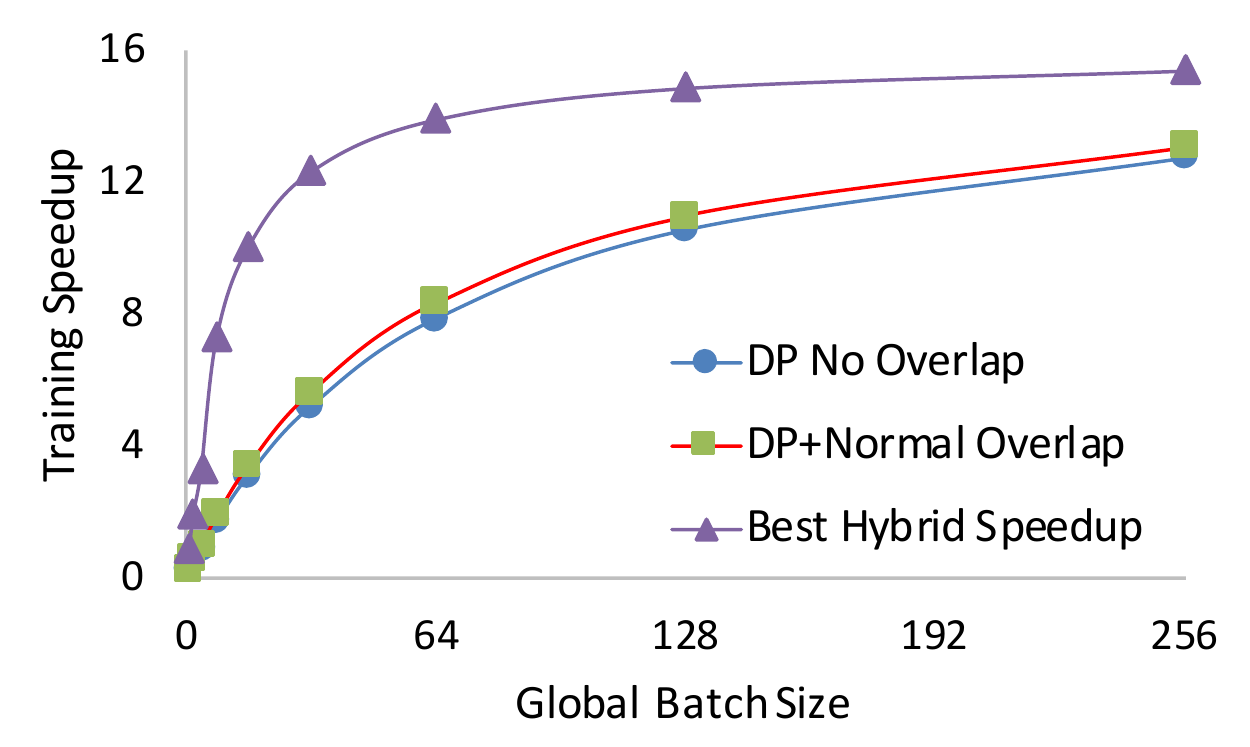}}
    \subfloat[XLNet-36 on config $B$]{\includegraphics[width=0.33\textwidth]{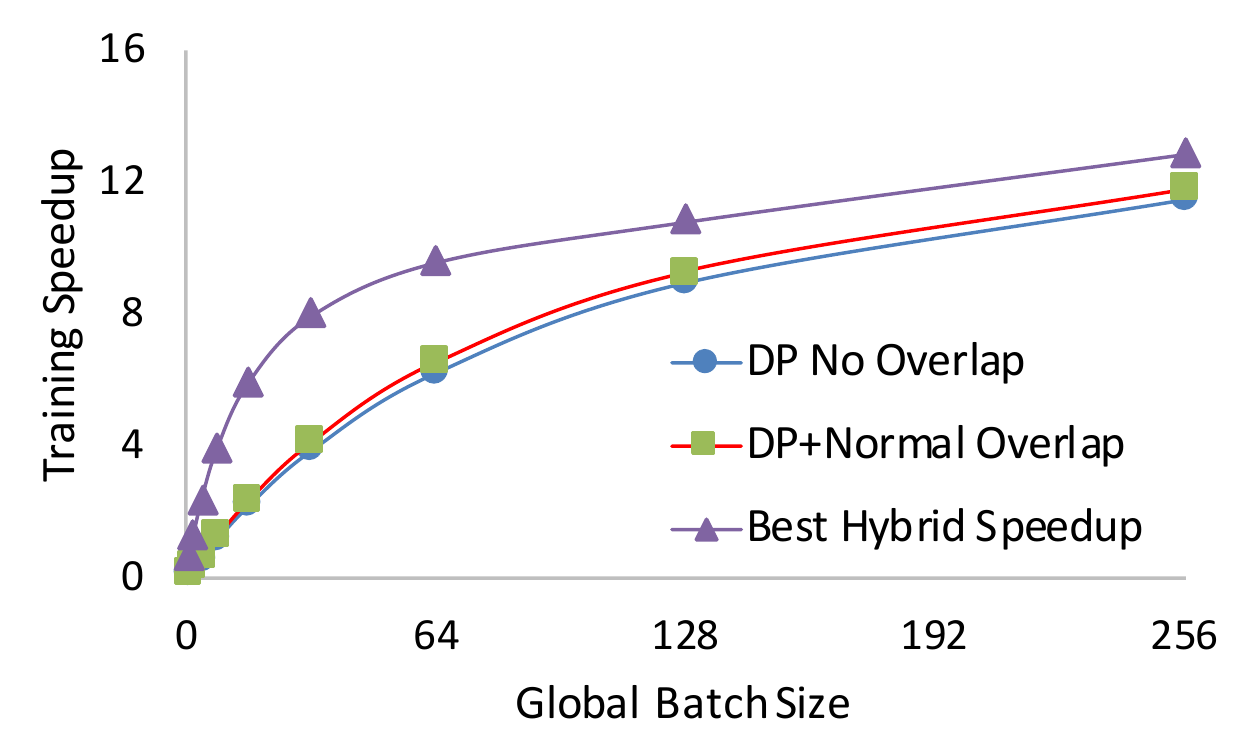}}
    \subfloat[XLNet-36 on config $C$]{\includegraphics[width=0.33\textwidth]{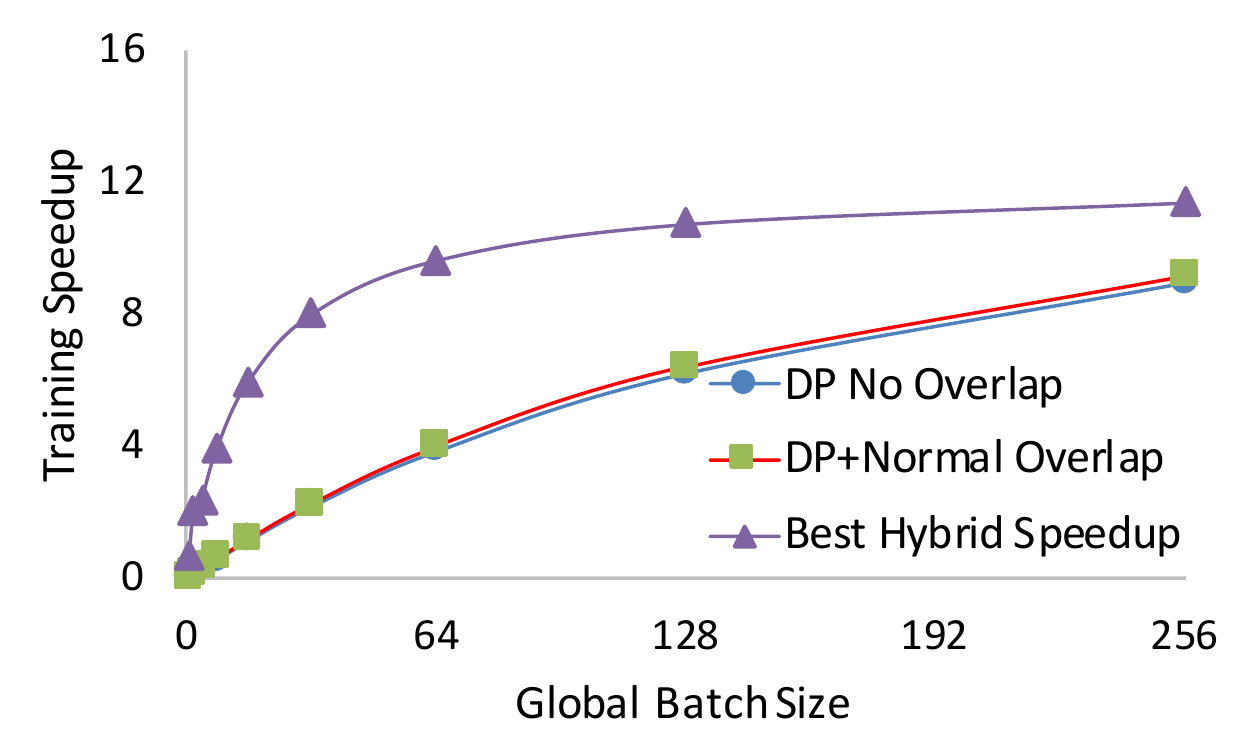}} \\
    \subfloat[AmoebaNet-36 on config $A$]{\includegraphics[width=0.33\textwidth]{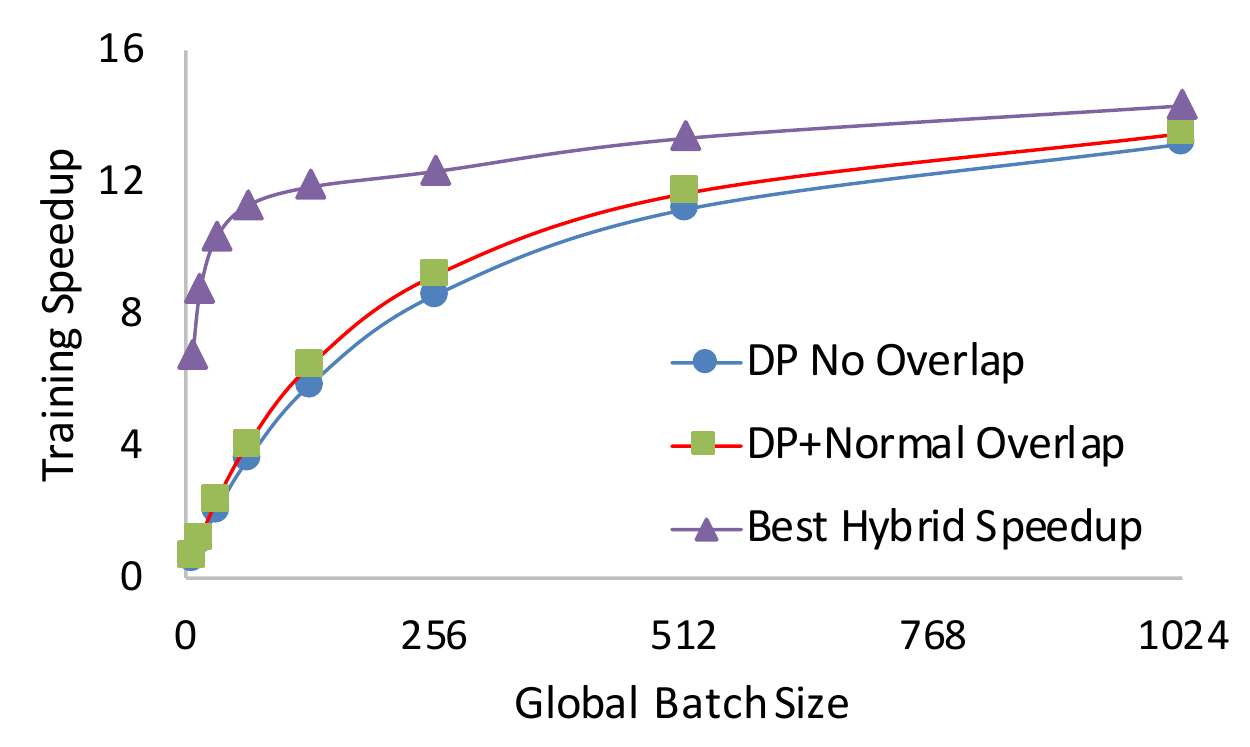}}
    \subfloat[AmoebaNet-36 on config $B$]{\includegraphics[width=0.33\textwidth]{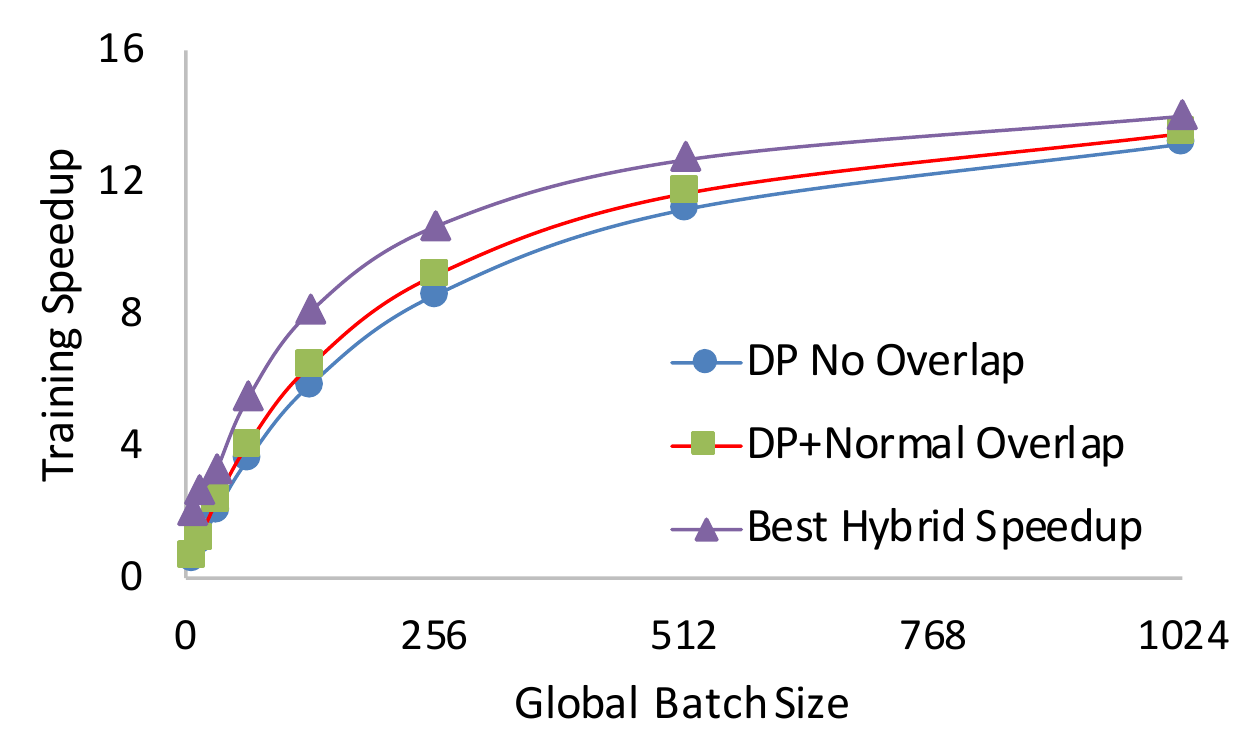}}
    \subfloat[AmoebaNet-36 on config $C$]{\includegraphics[width=0.33\textwidth]{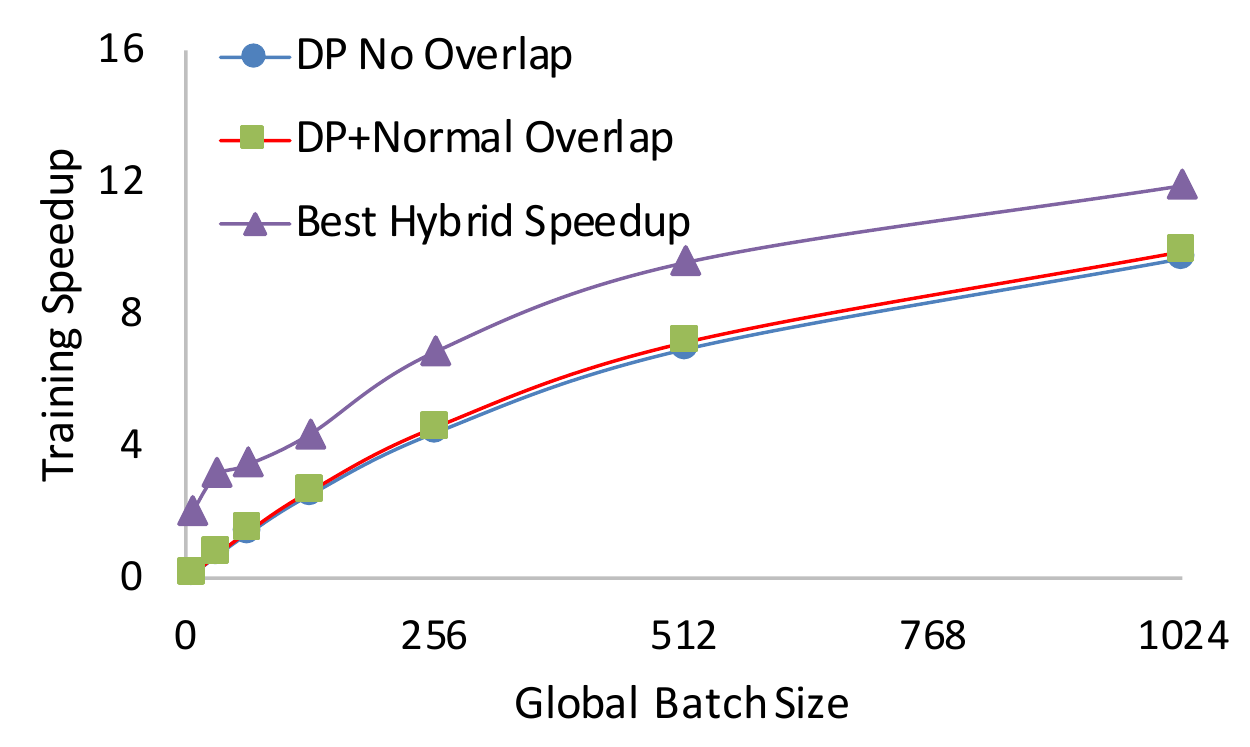}}
    \caption{
    Speedups on configurations with hierarchical/flat interconnects.
    }
    \label{fig:perf-speedup}
    % \vspace{0.2in}
\end{figure*}

\begin{table}[t]
    \caption{
     \emph{DAPPLE} planning results.
     % TBD: The planning results need be double checked by @rongyi
     % Note: ACR computation:
     %  (1) activation size x 2 / 800 MB (fw+bw, p2p comm speed!)
     %  (2) computation time: choose which stage? average? the slowest?
     %  (3) choosing which cut-surface? As the activation varies as GBS changes.
    }
    \label{tbl:results-overview}
    \begin{center}
    \begin{tabular}{lrrrr}
    \toprule
    \makecell[l]{Model\\(\emph{GBS})} & $\#Servers\times N_s$ & \makecell[c]{Output\\Plan} & \makecell[c]{Split\\Position} & $ACR$ \\
    \midrule
    \multirow{2}{*}{\makecell[l]{ResNet-50\\(2048)}}
    % & $1\times8$ (A) & DP & - & -\\
    & $2\times8$ (A) & DP & - & -\\
    & $16\times1$ (B) & DP & - & -\\
    & $16\times1$ (C) & DP & - & -\\
    \cline{1-5}
    % TBD: whether GBS=1024 is suitable for VGG19
    \multirow{2}{*}{\makecell[l]{VGG-19\\(2048)}}
    % & $1\times8$ (A) & DP & - & -\\
    & $2\times8$ (A) & DP & - & -\\
    & $16\times1$ (B) & DP & - & -\\
    & $16\times1$ (C) & $15:1$ & $13:6$ & 0.40\\
    \cline{1-5}
    % GNMT-32: batch_size=64, 452ms/step, activation_size=17.8MB
    \multirow{2}{*}{\makecell[l]{GNMT-16\\(1024)}}
    & $2\times8$ (A) & $8:8$ & $9:7$ & 0.10 \\
    & $16\times1$ (B) & $8:8$ & $9:7$ & 0.10 \\
    & $16\times1$ (C) & Straight & - & 3.75 \\
    \cline{1-5}
    \multirow{2}{*}{\makecell[l]{BERT-48\\(64)}}
    & $2\times8$ (A) & $8:8$ & $23:25$ & 0.06 \\
    & $16\times1$ (B) & Straight & - & 0.50 \\
    & $16\times1$ (C) & Straight & - & 1.25 \\
    \cline{1-5}
    \multirow{2}{*}{\makecell[l]{XLNet-36\\(128)}}
    & $2\times8$ (A) & $8:8$ & $18:18$ & 0.03 \\
    & $16\times1$ (B) & $8:8$ & $18:18$ & 0.03 \\ % when gbs <= 120: straight; else 2 stage
    & $16\times1$ (C) & Straight & - & 0.67 \\ % when gbs <= 96, else: 2 stage pipe
    \cline{1-5}
    % \multirow{3}{*}{\makecell[c]{Amoeba\\Net-18}} & $1\times8$ (A) & ? & ? & -\\
    % & $2\times8$ (B) & $8:8$ & 13:5 & 0.27 \\
    % & $16\times1$ (C) & $11:5$ & 13:5 & 0.27 \\
    % & $16\times1$ (D) & $8:8$ & ? & ? \\
    % \cline{1-5}
    \multirow{2}{*}{\makecell[l]{AmoebaNet-36\\(128)}}
    & $2\times8$ (A) & $8:8$ & $24:12$ & 0.18 \\
    & $16\times1$ (B) & $11:5$ & $27:9$ & 0.14 \\
    & $16\times1$ (C) & $11:5$ & $27:9$ & 0.35 \\
    \cline{1-5}
    \bottomrule
    \end{tabular}
    \end{center}
\end{table}

Table \ref{tbl:results-overview} summarizes \emph{DAPPLE} planning
results of five models in the three hardware environments,
where the total number of available devices are all fixed at $16$.
The first column also gives the global batch size ($GBS$) correspondingly.

We use three notations to explain the output plans.
(1) A plan of \emph{$P:Q$} indicates a two stage pipeline, with the first stage 
and the second stages replicated on $P$ and $Q$ devices, respectively.
For example, when $P=8$ and $Q=8$, we put each stage on one server, and 
replicate each stage on all 8 devices within the server(\emph{config-A}).
%When $P=15$ and $Q=1$, the first stage is replicated on $15$ GPUs while the second one on $1$ GPU.
Besides, for plans where $P>8$ or $Q>8$ (e.g., $15:1$) where some stages are replicated across servers,
it will most likely be chosen for configurations with flat interconnections such as 
\emph{Config-B} or \emph{Config-C}, since for \emph{Config-A} replicating one stage across servers
incurs additional inter-server communication overhead.
(2) A \emph{straight} plan denotes pipelines with no replication. 
(3) A \emph{DP} plan means the optimal strategy is data-parallel.
We treat \emph{DP} and \emph{straight} as special cases of general
\emph{DAPPLE} plans.

The \emph{Split Position} column of Table \ref{tbl:results-overview} shows the
stage partition point of each model for the corresponding pipeline plan.
The $ACR$ column of the table shows the $averaged$ ratio of cross-stage communication latency
(i.e. communication of both activations in FW and gradients in BW) and stage computation time.

In the case of single server of config $A$, there is
no relative low-speed inter-server connection,
the intra-server bandwidth is fast enough (up to $130GB/s$) to easily handle
the magnitude (up to $3.7GB$) of gradients communication of all benchmark models,
and we find all models prefer DP plan for this case.

\textbf{ResNet-50}. The best plan is consistently \emph{DP} for all three hardware
configurations. This is not surprising due to its relatively small model
size (100MB) yet large computation density. 
Even with low speed interconnects config $C$, \emph{DP} with notably gradients accumulation 
and computation/communication overlap outperforms the pipelined approach.

\textbf{VGG-19}. Best plans in config $A$ and $B$ are also \emph{DP}
(Fig. \ref{fig:perf-speedup} (a) and(b)), due to the moderate
model size (548MB), relatively fast interconnects (25 Gbps), and the overlapping 
in \emph{DP}. The weights and computation distributions of VGG19 are also
considered overlapping-friendly, since most of the weights are towards the end
of the model while computations are at the beginning, allowing gradients
aggregation to be overlapped during that computation-heavy phase. In the case of
low speed interconnects (config $C$), a $15:1$ pipelined
outperforms \emph{DP} (Fig. \ref{fig:perf-speedup} (c).
This is because most parameters in VGG-19 agglomerate 
in the last fully connected layer. A $15:1$ two-stage pipeline thus avoids most of the 
overheads of gradients synchronization due to replication 
(note we do not replicate the second stage).
In this case gradients synchronization overheads outweigh benefits of 
\emph{DP} with overlap.

\textbf{GNMT-16/BERT-48/XLNet-36}. All three models have uniform layer structures,
i.e., each layer has roughly the same scale of computations and parameters.
And the parameter scales of these models vary
from 1.2 GB up to 2.6 GB(Table \ref{tbl:models}).
In \emph{config-A} where all three models achieve low $ACR$ values (0.10, 0.06 and 0.03, respectively,
as shown in Table \ref{tbl:results-overview}), a two stage $8:8$ pipeline works best.
Unlike VGG-19, the three models' layers are relatively uniformly distributed, thus a symmetric,
evenly partitioning is more efficient.
In config $C$, a straight pipeline works best for all three models. In this config,
all devices have approximately the same workload. More importantly, no replication eliminates
gradients sync overheads for relatively large models (1.2-2.6 GB) on a slow network (10 Gbps).
The three models behave differently in config $B$.
BERT-48 prefers straight pipeline in config $B$, while GNMT-16 and XLNet-36 keep the same plan
results as shown in config $A$. This is because for fixed $16$ devices, $16$ and $48$ uniform layers
are more friendly for even partition compared to $36$ layers for flat interconnections.

\textbf{AmoebaNet-36}. For AmoebaNet-36, \emph{DP} is not available due to device memory limit. 
%of individual GPUs.
%For this case we profile the model on 2 devices where a 2-stage pipeline is applied, and then
%the profiling results are \emph{projected} to single device for comparative analysis, as shown in
%Fig. \ref{fig:perf-speedup}(m)(n)(o).
AmoebaNet-36 has more complex network patterns than other models we evaluated, and larger
ACR in config $A$ as well.
Thus, more successive forward micro-batches are needed to saturate the pipeline.
For all three configs, two-stage pipeline ($8:8$, $11:5$ and $11:5$, respectively) works best.

\subsection{Performance Analysis}

In this work, we measure \emph{training speed-up} as the ratio between the time executing all
micro-batches sequentially on a single device and the time executing all micro-batches in parallel by
all devices, with the same global batch size. 

Fig. \ref{fig:perf-speedup} shows training speed-ups for all models except ResNet-50 on config A, B and C.
For ResNet-50, the planning results are obvious and we simply present it in Table \ref{tbl:results-overview}.
For the other models, we compare training speed-ups of three different implementations:
(1) \textbf{Best Hybrid Speedup}, performance of the best hybrid plan of pipeline and data parallelism returned by \emph{DAPPLE} planner;
(2) \textbf{DP No Overlap}, performance of \emph{DP} with gradients accumulation but without computation/communication overlap;
(3) \textbf{DP Overlap}, performance of \emph{DP} with both gradients accumulation and intra-iteration 
computation/communication overlap between backward computation and gradients communication\cite{poseidon}.
% (4) \textbf{DP+P3}, performance of \emph{DP} with both intra and cross iteration overlap
% optimizations \cite{poseidon,jayarajan2019priority}.
%We consider all these \emph{DP} variants because we hope to understand how far 
%\emph{DAPPLE} can push beyond the most aggressive optimizations of \emph{DP}.

Overall analysis across these five models from Fig. \ref{fig:perf-speedup}, for fixed $GBS=128$,
we can find that the hybrid approaches from \emph{DAPPLE} outperform the \emph{DP} approach with best intra-batch overlapping
with averaged 1.71X/1.37/1.79X speedup for \emph{config-A}, \emph{config-B} and \emph{config-C}, respectively.
Specially, this speedup is up to 2.32X for GNMT-16 on \emph{config-C}.
Specific analysis for each model is given below.

\textbf{VGG-19}. %\emph{DP} works best in \emph{config-A} and \emph{config-B} (Fig. \ref{fig:perf-speedup} (a) and (b), ), while 
For VGG-19, about 70\% of model weights (about 400 MB) are in the last fully 
connected (fc) layer, while the activation size between any two 
adjacent layers gradually decreases from the first convolution layer to the last fc layer, 
varying dramatically from 384 MB to 3 MB for batch size 32.
Thus, the split between VGG-19's convolutional layers and fully-connected layers 
leads to very small activation (3MB), and only replicating all the convolutional layers 
other than fully-connected layers greatly reduces communication overhead in case of  
relatively slow interconnects (Fig. \ref{fig:perf-speedup} (c)).

\textbf{GNMT-16}.
GNMT-16 prefers a two-stage pipeline on hierarchical
network (config $A$) and flat network with relative high-speed connection (config $B$).
And the corresponding spit position is $9:7$ but not $8:8$, this is because
the per-layer workloads of encoder layer and decoder of GNMT
are unbalanced (approximately $1:1.45$), thus the split position of
\emph{DAPPLE} plan shifts one layer up into decoder
for pursuit of better system load-balance.
For low speed interconnection environments (config $C$), straight
pipeline ranks first when $GBS=1024$. Each device is assigned exactly one LSTM layers of GNMT, and
the $GBS$ is large enough to fill the 16-stage pipeline.

\textbf{BERT-48/XLNet-36}. The best \emph{DAPPLE} plan outperforms all \emph{DP} variants 
for both models (Fig. \ref{fig:perf-speedup} (g) to (l)) in all configurations.
%However, the advantage margin over the most aggressive \textbf{DP+P3} is different.
Compared to XLNet, the memory requirement for
BERT is much smaller and thus allows more micro-batches on a single device.
More computation per-step implies more backward computation
time can be leveraged for overlapping communication overhead.
As for config $B$ and $C$, the slower the network is(from 25 Gbps to 10 Gbps), the higher 
the advantage of our approach has over \emph{DP} variants. This is because the cross stage communication 
for both models is negligible with respect to gradients communication and the pipelined approach is more tolerant of 
slow network than \emph{DP}.

\textbf{AmoebaNet-36}. 
The \emph{DAPPLE} plan works best in all three configurations when \emph{GBS}
is fixed to $128$.
Unlike BERT-48 and XLNet-36, AmoebaNet has non uniform distributions of per layer parameters and computation density.
The last third part of the model holds
73\% of all parameters, and the per-layer computation time gradually increases for large layer id and
the overall maximum increase is within 40\%.
As \emph{DAPPLE planner} seeks for load-balanced staging scheme while considering
the \emph{allreduce} overhead across replicated stages, the \emph{split
  positions} of pipelined approach for AmoebaNet-36 will obviously tilt to
larger layer ID for better system efficiency.
Take config $A$ as an example, a two-stage pipeline is chosen and each stage is replicated over a separate
server with $8$ devices each. For this case a total of $36$ normal cells layers are divided into $2$ parts,
namely $24:12$, where the per-stage computation time ratio and \emph{allreduce} time
%(i.e., gradients synchronization between replicated stages each)
ratio of $stage_0$ and $stage_1$ is $1.57:1$ and $1:2.74$, respectively.
For lower bandwidth network configurations (config $B$ and $C$), the \emph{split position}
keeps shrinking to larger layer ID, because the allreduce \emph{comm} overheads turn out to
be the dominant factor with the absent of high-speed NVLink bandwidth.

\subsection{Scheduling Policy}
\label{sec:scheduling-policy}

As discussed in Section \ref{section:micro-batch-schedule}, the number of successive
forward micro-batches ($K_i$ for stage $i$) scheduled in the warm up phase is an important factor to 
pipeline efficiency. We implement two policies, $P_A$ and $P_B$, referring to smaller and
larger $K_i$ numbers, respectively.
Table \ref{tbl:shcedule-policy} shows the normalized speedups
for four benchmark models on hierarchical interconnects(config $A$),
where all models' stage partition and replication schemes are consistent with the planning results of 2 servers of config $A$
as shown in Table \ref{tbl:results-overview}.

For VGG-19 and GNMT-16 (as well as AmoebaNet-36, which is not given in this figure yet),
where the $ACR$ ratio is relative high (0.16, 0.10, 0.18, respectively),
there exists notable performance difference between these two
policies (10\%, 31\% improvement from $P_A$ to $P_B$, respectively).
Hence we choose a larger $K_i$ to maximize pipeline efficiency.
For the other models (BERT-48, XLNet-36), whose $ACRs$ are very small
(0.06, 0.03, respectively), the cross stage communication overhead is
negligible compared to intra-stage computation time, leading to little performance difference.
In this case, we prefer a smaller $K_i$ to conserve memory consumption.

%\subsection{Compared to GPipe and PipeDream}

\begin{table}[t]
    \caption{
        \emph{DAPPLE} vs. GPipe on BERT-48 with $2$-stage pipeline when keeping micro-batch size fixed to $2$ on \emph{Config-B}.
        \emph{RC} is short for re-computation.
        }
    \label{tbl:compared-to-gpipe}
    \begin{center}
    \begin{tabular}{cccc}
    \toprule
    % Config & \makecell[c]{GPU(s) per\\server($N_s$)} & \makecell[c]{Intra-server\\connnections} &\makecell[c]{Inter-server\\connections} \\
    Config & \makecell[c]{\# of micro \\batch ($M$)} & \makecell[c]{Throughput \\(samples/sec)} & \makecell[c]{Average Peak\\ Memory (GB)} \\
    \midrule
    \multirow{2}{*}{GPipe}
    & 2 & 5.10 & 12.1 \\
    & 3 & -- & $OOM$ \\
    \cline{1-4}
    \multirow{3}{*}{GPipe + RC}
    & 2 & 4.00 & 9.9 \\
    & 5 & 5.53 & 13.2 \\
    & 8 & -- & $OOM$ \\  % @FSQ: need to be confirmed.
    \cline{1-4}
    \multirow{3}{*}{\emph{DAPPLE}}
    & 2 & 5.10 & 10.6 \\
    & 8 & 7.60 & 10.6 \\
    & 16 & 8.18 & 10.6 \\
    \cline{1-4}
    \multirow{3}{*}{\emph{DAPPLE} + RC}
    & 2 & 4.24 & 8.5 \\
    & 8 & 6.23 & 8.5 \\
    & 16 & 6.77 & 8.5 \\
    \cline{1-4}
    \bottomrule
    \end{tabular}
    \end{center}
\end{table}

\subsection{{Comparison with GPipe}}
% \label{sec:compared-to-gpipe}
% Here we discuss \emph{DAPPLE} results in the context of related research on pipeline training.

% \textbf{GPipe}\cite{huang2019gpipe}.

Table \ref{tbl:compared-to-gpipe} shows the performance comparisons
with GPipe. We focus on the throughput and peak memory usage on BERT-48 with a 2-stage pipeline on \emph{Config-B}. To align with GPipe, we adopt the same re-computation strategy which stores activations only at the partition boundaries during forward. Note that all the pipeline latency optimizations in \emph{DAPPLE} give equivalent gradients for training when keeping global batch size fixed and thus convergence is safely preserved and will not be further analysed here.

When applying \emph{re-computation}, both \emph{DAPPLE} and GPipe save about 19\% 
averaged peak memory at the expense of 20\% on throughput when keeping $M=2$ fixed.
% , which is consistent with previous reports \cite{gpipeTalk19}.

When both without \emph{re-computation}, \emph{DAPPLE} gets $1.6\times$ higher throughput with $M=16$, and consumes $0.88\times$ averaged peak memory compared to GPipe, which only supports up to 2 micro-batches. The speedup is mainly because higher $M$ leads to lower proportion of $bubbles$. Note \emph{DAPPLE} allows more micro-batches as the peak memory requirement is independent of $M$ due to \emph{early backward scheduling}.

The combination of \emph{DAPPLE} scheduler and re-computation allows a 
further exploitation in memory usage. Compared with baseline GPipe (without re-computation), \emph{DAPPLE + RC} achieves $0.70\times$ memory consumption when $M=16$, which allows us to handle larger micro-batch size or larger model.

\subsection{Comparison with PipeDream}
\label{sec:comparison-with-pipedream}
% \textbf{PipeDream}\cite{harlap2018pipedream}.

% generated table
\begin{table}[]
\centering
\caption{ Strategy Comparison between DAPPLE and PipeDream, 
in the form of (start layer, end layer)@[GPU IDs].}
\label{tbl:strategy-comparison}
\begin{tabular}{@{}lll@{}}
\toprule
Model (GBS)   & DAPPLE                                                                                      & PipeDream \\ \midrule
VGG19 (1024) &
  \begin{tabular}[c]{@{}l@{}} (0, 16) @ {[}G0 - G13{]}\\ (17, 25) @ {[}G14, G15{]} \end{tabular} &
  \begin{tabular}[c]{@{}l@{}} (0, 11) @ {[}G0 - G7{]} \\ (11, 17) @ {[}G8 - G13{]} \\ (17, 19) @ G14\\ (19, 25) @ G15 \end{tabular} \\ \midrule
AmoebaNet-36 (128) & \begin{tabular}[c]{@{}l@{}} (0, 30) @ {[}G0 - G7{]}\\ (31, 43) @ {[}G8 - G15{]}\end{tabular} & straight  \\ \midrule
BERT Large (128) &
  \begin{tabular}[c]{@{}l@{}}(0, 13) @ {[}G0 - G7{]}\\ (14, 26) @ {[}G8 - G15{]}\end{tabular} &
  \begin{tabular}[c]{@{}l@{}} (0, 4) @ {[}G0, G1{]}\\ (4, 13) @ {[}G2 - G7{]}\\ (13, 16) @ {[}G8, G9{]}\\ (16, 19) @ {[}G10, G11{]}\\ 
(19, 22) @ {[}G12, G13{]}\\ (22, 26) @ {[}G14, G15{]}\end{tabular} \\ \midrule
XLNet-36 (128)     & \begin{tabular}[c]{@{}l@{}} (0, 22) @ {[}G0 - G7{]}\\ (23, 41) @ {[}G8 - G15{]} \end{tabular} & straight  \\ \bottomrule
\end{tabular}
\end{table}

\begin{figure}[!tbp]
  \centering
  \includegraphics[width=\linewidth]{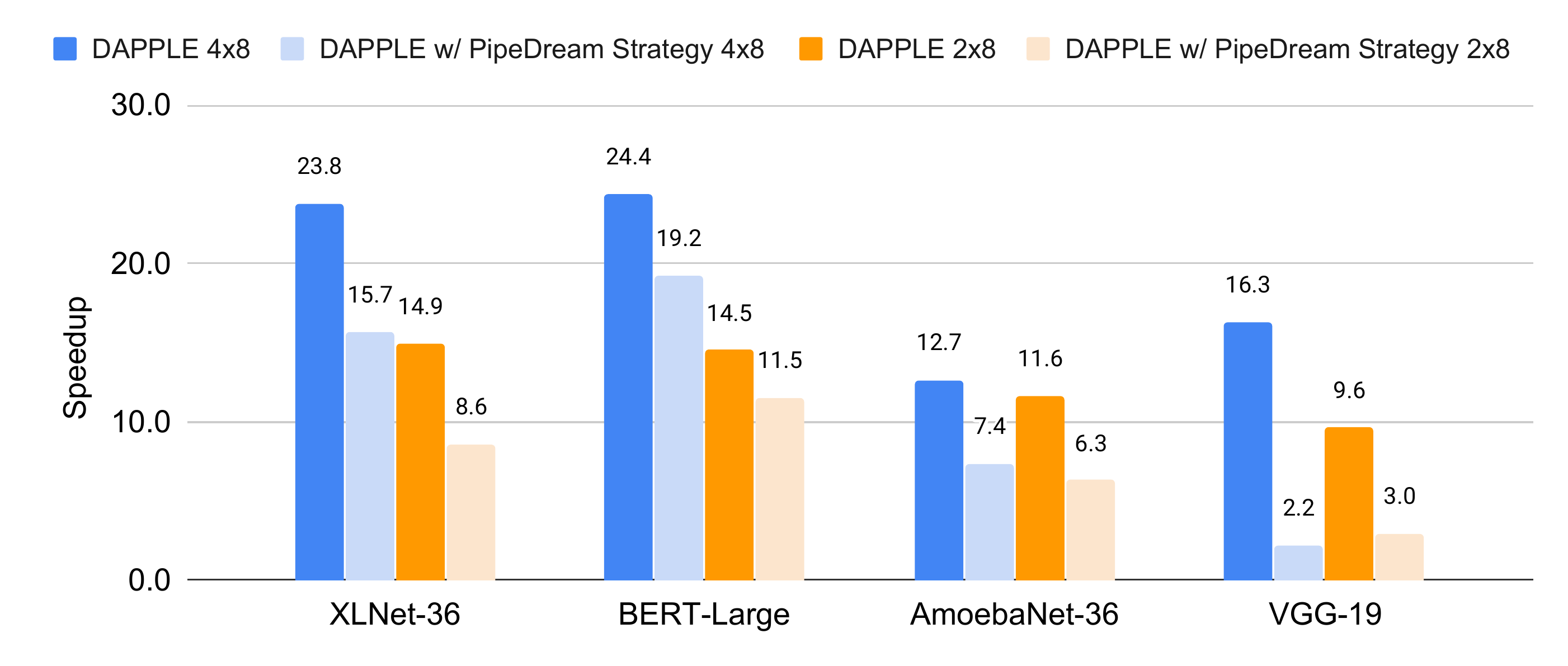}
  \caption{
    Performance comparison with PipeDream.
  }
  \label{fig:HPGO-vs-PipeDream}
\end{figure}

We compare the results of our planner with those of PipeDream's under the 
synchronous training scenarios. We use the same configurations for both planners 
(e.g. same device topology, same interconnect and same profiling
data), and evaluate both planners with \emph{DAPPLE} Runtime. Table
\ref{tbl:strategy-comparison} shows the strategy results under a two-machine cluster of \emph{config-A}.
Fig. \ref{fig:HPGO-vs-PipeDream} shows the
performance results for the strategies running in both $2\times8$ and $4\times8$ configurations.

As shown in Fig. \ref{fig:HPGO-vs-PipeDream}, in terms of $speedup$ relative to to data parallelism, our strategies consistently outperform those generated by PipeDream's planner by up to $3.23\times$ speedup under synchronous training scenarios, 
thanks to the advantages detailed in Section \ref{sec:contribution-over-previous-work}.

\subsection{Strong Scaling}
\begin{figure*}[!htb]
    \centering
    \subfloat[GNMT-16($GBS=2048$)]{\includegraphics[width=0.25\textwidth]{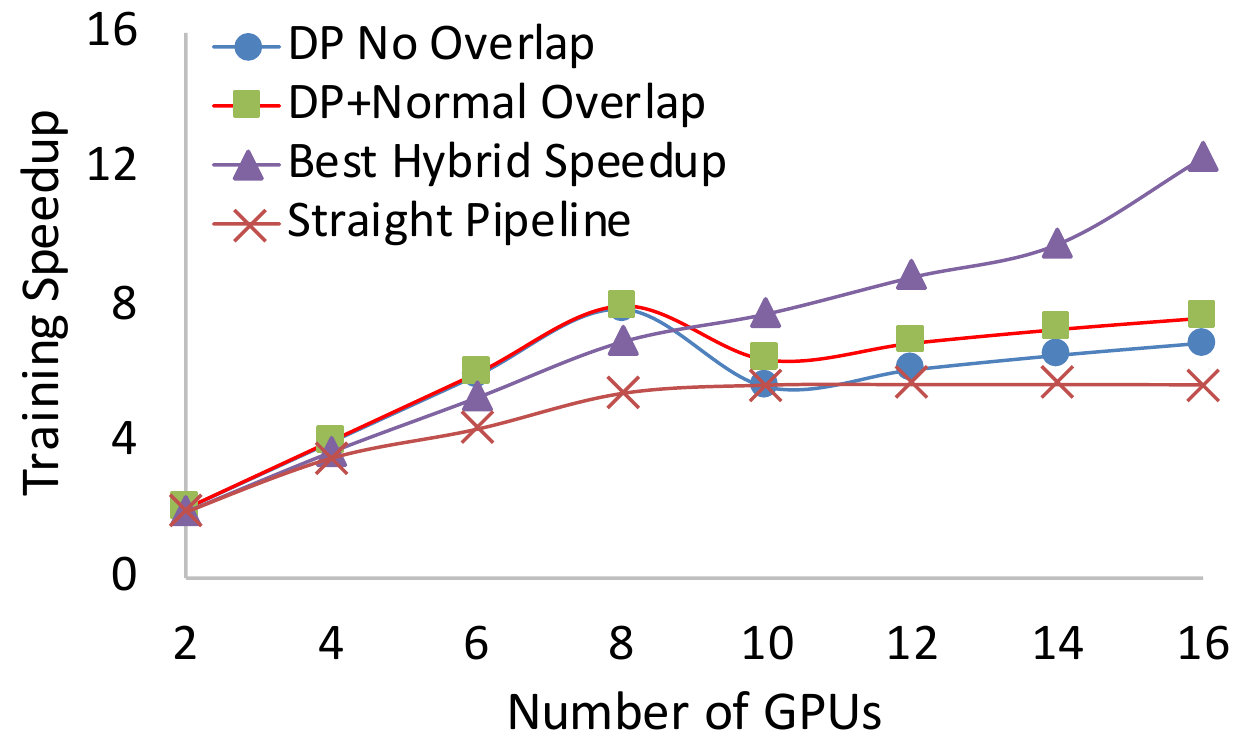}}
    \subfloat[BERT-48($GBS=128$)]{\includegraphics[width=0.25\textwidth]{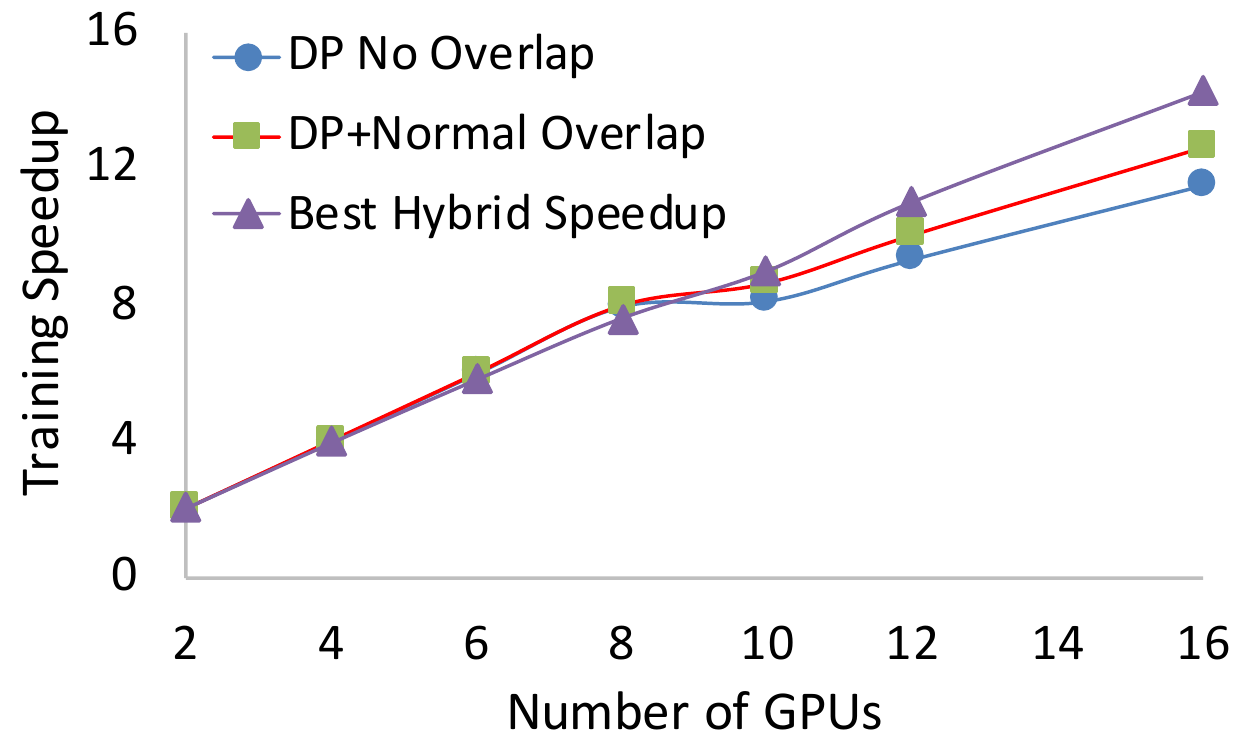}}
    \subfloat[XLNet-36($GBS=128$)]{\includegraphics[width=0.25\textwidth]{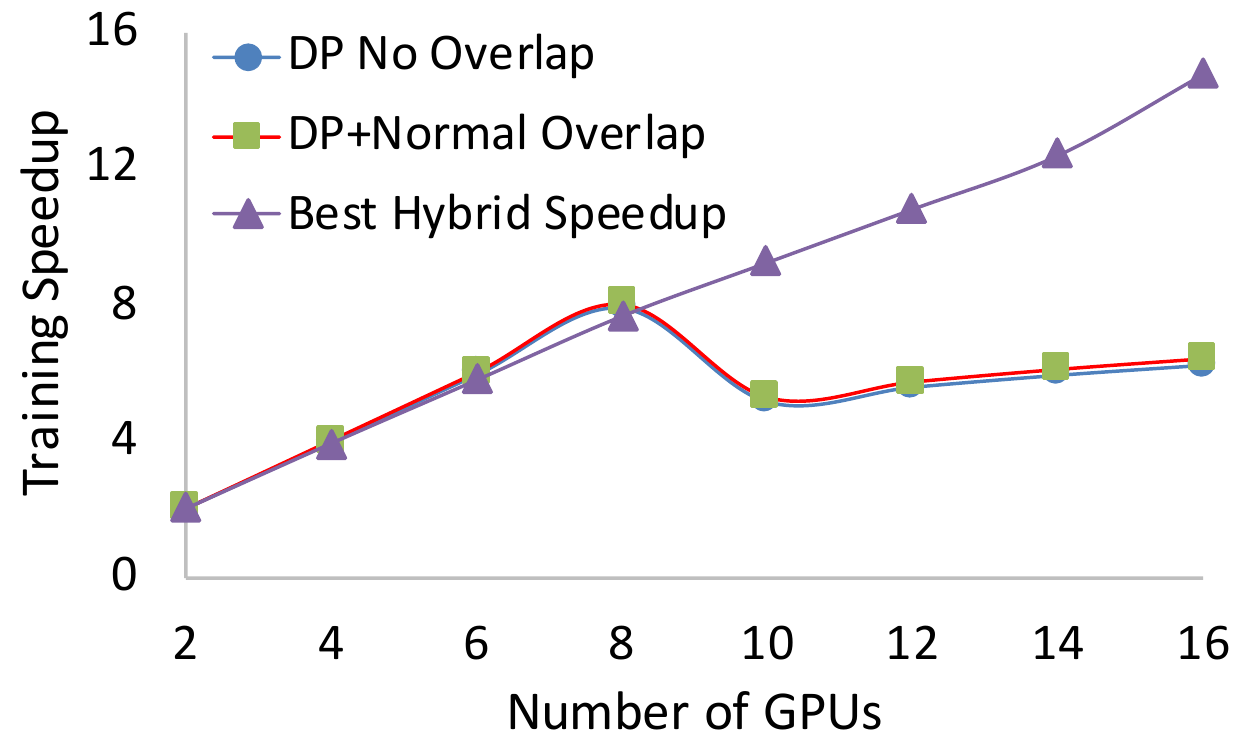}}
    \subfloat[AmoebaNet-36($GBS=256$)]{\includegraphics[width=0.25\textwidth]{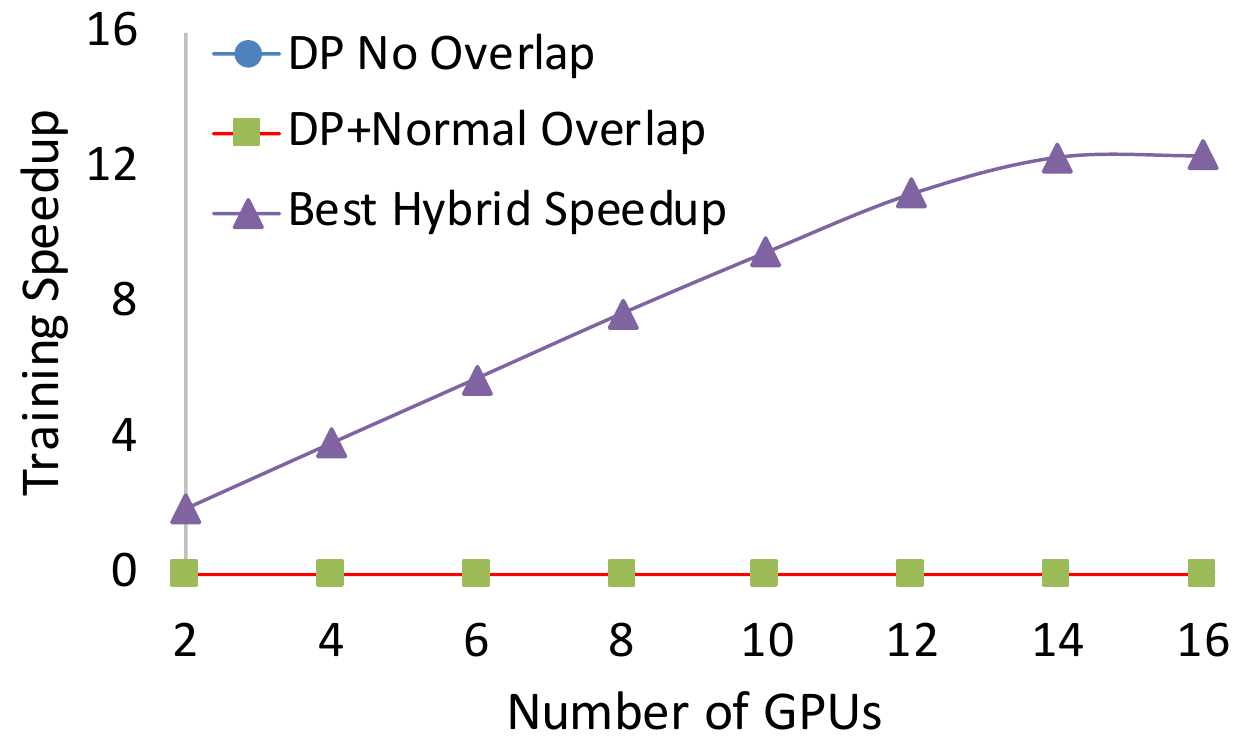}} \\
    \caption{
    Speedup with fixed $GBS$ in \emph{config-A}.
    }
    \label{fig:q-to-gbs}
    \vspace{0.2in}
\end{figure*}

Fig. \ref{fig:q-to-gbs} shows training speed-ups for four models. The number of GPUs ranges from 2 to 16. We use fixed but different global batch size for each model and apply config $A$. For AmoebaNet-36 when $GBS=256$ (Fig. \ref{fig:q-to-gbs}(d)), both \emph{DP} approaches achieve NaN as the model size is too large to fit memory of single V100. For all these four models, we observe better scalability of \emph{DAPPLE} over \emph{DP} variants. Scalability weakens on all \emph{DP} variants when the number of GPUs increases from 8 to 10, where gradients synchronization performance drops substantially due to slow Ethernet bandwidth. The performance of $DAPPLE$ approach scales smoothly due to the rather small magnitude of cross-stage activations as compared with weights(Table \ref{tbl:comm-vol}), which is insensitive to the relatively low-speed inter-server communications(25Gbps). In general, for hierarchical interconnects, the lower the cross-machine bandwidth, the more obvious the advantage $DAPPLE$ approach as compared with $DP$.

\subsection{Weak Scaling}
    Table \ref{tbl:scalability} shows the maximum model size that \emph{DAPPLE}
    supports under reasonable input size with re-computation.
    We scale the model by varying the numbers of layers.
    We are able to scale BERT to 5.5B on 8 V100s with NVLink. 
    There is a slight reduction in average GPU utilization due to more bubbles introduced by 
    longer pipeline. In this case, the maximum model size scales linearly due to
    the balanced distribution of model params over encoder layers in BERT.
\begin{table}[t]
    \caption{
        Maximum model size of BERT supported by \emph{DAPPLE}
    + re-computation on V100 (16GB each) on \emph{config-A}. BERT-L: BERT model with $L$ encoder layers. Each model parameter needs 16 bytes since we applied Adam optimizer. }
    
    \label{tbl:scalability}
    \begin{center}
    \begin{tabular}{lcccc}
    \toprule
    Config & BERT-L & \makecell[c]{\# of Model\\Params} & \makecell[c]{Total Model\\Params Mem} & \makecell[c]{Avg. GPU\\ Util}\\
    \midrule
    % A & 8x V100 & NVLink & N/A \\
    Native-1    & 48   & 640M   & 10.2GB  & 93\% \\
    Pipeline-2  & 106  & 1.4B   & 21.9GB  & 89\% \\
    Pipeline-4  & 215  & 2.7B   & 43.8GB  & 89\% \\
    Pipeline-8  & 428  & 5.5B   & 88.2GB  & 87\% \\ % ??
    
    \bottomrule
    \end{tabular}
    \end{center}
\end{table}

\section{Related Work}
\label{section:relatedwork}

Large DNN models are increasingly computational intensive.
It is a common practice to parallelize training
by leveraging multiple GPUs\cite{dean2012large,pal2019optimizing,jia2018exploring,geng2019horizontal}.
Data parallelism, model parallelism and pipeline parallelism are common approaches for distributed training
of DNN models.
% Note we discuss pipeline parallelism separately from model parallelism.

\textbf{Data Parallelism}\cite{krizhevsky2014one} .
% Data Parallelism\cite{krizhevsky2014one} is a common approach for
% traditional distributed machine learning.
%When data parallelism is applied, each GPU keeps a replica of
%the entire network, which results in excessive inter-GPU communication
%overheads for synchronizing the gradients each step and limits its
%scalability.
Some prior studies \cite{BaiduAllReduce2018, jia2018highly,
arivazhagan2019massively, BytePS2019, sergeev2018horovod, you2019large} focus on reducing the communication
overheads for data parallelism.
% For instance, ring-based allreduce\cite{BaiduAllReduce2018,jia2018highly}
% and BytePS\cite{BytePS2019} can achieve high
% system scalability with constant communication costs.
%When using ring-based allreduce to exchange $K$ bytes of data with
%$N$ GPUs, each machine sends and receives $2(N-1)K/N$ bytes of data,
%which is independent of the number of GPUs.
% Horovod\cite{sergeev2018horovod} proposes gradient fusion strategy
% to address small message transmission problem for ring-based
% allreduce in large scale clusters.
As a commonly used performance optimization method, gradients accumulation\cite{le2018involving,GA-Torch,GA-TF}
offers an effective approach to reduce communication-to-computation ratio.
%and thus improve device utilization.
Another complementary approach is computation and communication overlap,
with promising results reported in some CNN benchmarks\cite{poseidon,jayarajan2019priority}.
%Le et al. \cite{le2018involving} leverage the host for performing
%gradient accumulation to accelerate training on GPUs.
%Jayarajan et al. \cite{jayarajan2019priority} propose P3, which
%improves the training performance by better utilizing the
%available network bandwidth.
%However, the limited GPU memory fails the data-parallel solutions
%as the model grows larger.

\textbf{Model Parallelism}. Model Parallelism\cite{jia2018beyond} partitions DNN models among GPUs
to mitigate communication overhead and memory bottlenecks
for distributed training
\cite{huo2018decoupled,pal2019optimizing,dean2012large,geng2019rima,harlap2018pipedream,huang2019gpipe,chen2018efficient,dryden2019channel}.
This paper focuses on model partition between layers, namely, pipeline parallelism.

%The pipe-based model parallelism can benefit from:
%1) overcoming the single node's GPU memory limitation
%through partitioning large model and distributing to each device.
%2) reducing communication overhead compared to data parallel,
%where only intermediate outputs (and corresponding gradients) of
%the boundary layers needs to transmit to its neighbours.
%However, this approach suffers from low resource utilization as
%only one device is active in the execution of pipeline workflow.
\textbf{Pipeline parallelism}. Pipeline Parallelism\cite{harlap2018pipedream,zhan2019pipe,huang2019gpipe,geng2019elasticpipe,yang2019pipemare}
has been recently proposed to train DNN in a pipelined manner.
%This approach achieves better overlap of communication and computation
%with each other, as communication and computation are executed in
%a finner granularity through the pipeline workflow.
GPipe\cite{huang2019gpipe} explores synchronous pipeline approach to train large models with 
limited GPU memory.
PipeDream\cite{harlap2018pipedream} explores the hybrid approach of data and pipeline parallelism
for asynchronous training.
\cite{guan2019xpipe,geng2019elasticpipe,chen2018efficient}
make further optimization based on PipeDream.
Pal et al. \cite{pal2019optimizing} evaluated the hybrid approach without thorough study.
Some researchers have been seeking for the optimal placement strategy
to assign operations in a DNN to different devices\cite{mirhoseini2017device, gu2019tiresias, wang2019supporting}
to further improve system efficiency.

\section{Conclusion}
\label{section:conclusion}

In this paper, we propose \emph{DAPPLE} framework for pipelined training of large DNN models.
\emph{DAPPLE} addresses the need for synchronous pipelined training and advances current 
state-of-the-art by novel pipeline planning and micro-batch scheduling approaches. 
On one hand, \emph{DAPPLE planner} module determines an optimal parallelization strategy given model structure 
and hardware configurations. It considers pipeline partition, replication and placement, 
and generates a high-performance hybrid data/pipeline parallel strategy.
On the other hand, \emph{DAPPLE scheduler} module is capable of simultaneously achieving optimal training efficiency 
and moderate memory consumption, without storing multiple versions of parameters and getting rid of the strong demand of re-computation which hurts system efficiency at the same time.
Experiments show that \emph{DAPPLE planner} consistently outperforms strategies generated by PipeDream‘s planner by up to $3.23\times$ speedup under synchronous training scenarios, and \emph{DAPPLE scheduler} outperforms GPipe by $1.6\times$ speedup of training throughput and saves 12\% of memory consumption at the same time.

% Experimental results on a set of large image classification, machine translation and
% language models show promising results (up to 2.32X compare to best \emph{DP} baseline). 
% \emph{DAPPLE} is open-source and will be made available to the public.
%
%In future work, we will extend \emph{DAPPLE} to explore design spaces of more model
%parallel optimizations, and validate the approach on more application domains.
%Text of paper \ldots
%Given any model structure and network specification (isomorphic/heterogeneous),
%DAPPLE produces a distribution strategy (data-parallel/straight-pipeline/mixture)
%which achieves the highest average GPU efficiency according to corresponding workload.
%We highlight three key attributes of DAPPLE:
%\begin{enumerate}
%    \item Efficiency: Through better use of heterogeneous network, DAPPLE
%    achieves average xxx speedup compared to any up-to-data synchronous
%    pipeline approach.
%    \item Flexibility: DAPPLE supports any deep network that can be
%    represented as a sequence of layers.
%    \item Reliability: DAPPLE utilizes synchronous gradient descent and
%    guarantees consistent training regardless of the number of partitions.
%    Besides, heuristic rules added in the Planner system makes sure the final
%    partition outputs are strong practical guidance.
%\end{enumerate}


\bibliographystyle{IEEEtran}
% \bibliography{paper.bib}

% \input{appendix/appendix}

\end{document}